\documentclass{JFM-FLM_Au}

\usepackage{xcolor}
\usepackage{soul}
\sethlcolor{yellow}

\usepackage{graphicx}
\usepackage{subcaption}
\usepackage{overpic}
\usepackage{float}
\usepackage{caption}
\usepackage{booktabs}
\usepackage{multirow}
\usepackage{makecell}
\usepackage{dcolumn}

\lefttitle{An-Xiao Han, Peng-Yu Duan, Ming-Ze Ma and Xi Chen}
\righttitle{Journal of Fluid Mechanics}

\title{Subharmonic instability of large-scale wavy structures in two-dimensional channels}

\author{An-Xiao Han\aff{1}, Peng-Yu Duan\aff{1}, Ming-Ze Ma\aff{1} \and Xi Chen\aff{1}}

\affiliation{Key Laboratory of Fluid Mechanics of Ministry of Education, Beihang University (Beijing University of Aeronautics and Astronautics), Beijing, 100191, PR China}

\corresau{Xi Chen, \email{chenxi97@outlook.com}}

\begin{document}
\maketitle

\begin{abstract}
A particular interest on two-dimensional turbulence is the inverse energy cascade from small to large sales, which leads to an energy condensation accompanied by the formation of large-scale vortical structures. Indeed, such a phenomenon is  observed in the two-dimensional channel (2DCH) with large Reynolds numbers, where prominent large-scale wavy structures play a central role in the momentum and energy transfer across the inhomogeneous wall-normal direction \citep{Falkovich2018}. Yet, the instability of these wavy structures remains poorly understood, and it is unknown whether they have the capacity to  generate turbulence. To address this, we first conduct the direct numerical simulation (DNS) of Navier-Stokes equations for 2DCH, then extract the large-scale wavy structures through the singular value decomposition, and finally perform a Floquet-based secondary instability analysis. Two bulk Reynolds numbers are examined in particular, i.e. $Re = 3000$ and $Re = 200000$, which lie on opposite sides of the transitional regime near $Re \approx 10000$ and cover the previously reported simulation domain. At $Re = 3000$, the large-scale wavy structure is found to be linearly stable, consistent with the laminar-like DNS flow field. However, at $Re = 200000$, a subharmonic torsional mode is identified, which leads to a definite growth rate ($\lambda_r = 0.18$) for the wavy structures with a half wave-length shift. Temporal reconstruction shows that this unstable mode deforms and splits into multiple wave trains and evolves in the opposite phase. Compared to the TS (Tollmien-Schlichting) wave of laminar flow, the subharmonic mode found here offers a novel understanding for the generation of turbulence in larger Reynolds number two-dimensional channels.
\end{abstract}



\section{\label{sec:level1}Introduction}

Turbulence has traditionally been characterized within the framework of three-dimensional (3D) flows, where the canonical Richardson–Kolmogorov energy cascade describes the transfer of kinetic energy from large scales to small scales until its eventual viscous dissipation.
However, under conditions of strong geometric confinement where the flow is essentially two-dimensional (2D), turbulence exhibits a fundamentally different physical paradigm. 
The theoretical foundation of 2D turbulence, pioneered by \citet{Kraichnan1967} and \citet{Batchelor1969}, predicts a dual cascade in a statistically steady state: a forward enstrophy cascade to small scales and an inverse energy cascade toward progressively larger scales. 
This insight led to the Kraichnan-Batchelor theory, which identifies a $k^{-5/3}$ spectrum for the inverse energy cascade and a $k^{-3}$ spectrum for the forward enstrophy cascade.

Classical theories of 2D turbulence often employ periodic domains to neglect boundary effects \citep{Kraichnan1967, Batchelor1969, kraichnan1980,boffetta2012} and focus on the energy transport and dissipation between different scales. However, boundary effects are critical in practical configurations.
For example, dynamics in 2D channel flows have been explored in gravity-driven soap films \citep{Kellay1995, Kellay2002}, where subsequent studies \citep{Tran2010, Vilquin2021} established a link between energy flux and skin friction.
Recently, \cite{chen2024} investigated 2D channels and identified scaling laws of turbulence fluctuations distinct from those of 3D flows, notably predicting that near-wall fluctuations scale as the $1/3$ power of the Reynolds number such that they diverge as $Re \to \infty$. This behavior is linked to the transport induced by large-scale wavy structures arising from the inverse energy cascade. The statistical properties of such inverse-cascade flows have been characterized by exact third-order structure functions \citep{Xie2018} and the bifractal nature of circulation \citep{Zhu2023}, while the Bolgiano-Obukhov scaling in two-dimensional convection further illuminates the cascade mechanism \citep{Xie2022}. Despite ongoing interest in the mean velocity \citep{Lvov2009, Samanta2014, Markeviciute2021} and vorticity profiles \citep{Falkovich2018} of 2D channels, the stability of the large-scale structures inherent to these flows remains poorly understood. It is unknown whether these large-scale coherent structures are stable or not and how turbulence is generated in a 2D channel configuration, which motivates the present study. 

For classical parallel shear flow, the transition from a laminar to a turbulent state typically involves the following stages \citep{1}: receptivity stage, linear growth stage (mainly the linear development of Tollmien-Schlichting waves \citep{2, 3}), nonlinear growth stage \citep{4}, secondary instability stage \citep{5}, and eventually the breakdown stage. 
For subcritical transition, the disturbance evolution enters a nonlinear growth stage after initial linear decay. In this stage, finite-amplitude two-dimensional waves develop and eventually reach a saturated state \citep{12,13}. Weak nonlinear theory \citep{6,7,8} and bifurcation theory \citep{9,10} are commonly used to analyze this stage. In some cases, these nonlinear states can undergo further bifurcations into more complex motions \citep{14}, such as vortex rings or chaos, corresponding to the so-called ``exact coherent structures" or nonlinear invariant solutions of the Navier-Stokes equations \citep{15}.
Recent theoretical advances in boundary-layer stability, such as the local scattering theory of \citet{Wu2016} and the asymptotic analyses of \citet{Dong2021, Dong2022}, have provided powerful frameworks for understanding instability mechanisms in complex flows, while global stability analyses of streamwise vortices \citep{Chen2019, Chen2022} have revealed the critical role of large-scale coherent structures in triggering transition. The present study adopts a Floquet-based secondary instability analysis to investigate the stability of large-scale wavy structures in two-dimensional channels, building upon these theoretical foundations.

For two-dimensional flows, substantial efforts have been devoted to understand the flow instability and transition process. \cite{sobey1986} mapped the complex bifurcation landscape, revealing how the base flow gives way to multiple stable asymmetric states. \cite{21} provided a detailed account of the route to chaos in this system, identifying isolated turbulent patches that prefigure full transition. Further studies have enriched this framework by examining specific mechanisms. \cite{kerswell2018} identified a ``boundary inflow instability'' that can disrupt even linearly stable flows, while \cite{Falkovich2018} highlighted a fundamental difference between flow types, showing that turbulence in pressure-driven channels saturates into a large-scale traveling wave. The understanding of coherent structures in two-dimensional turbulence has also evolved significantly. Early work by \cite{jimenez1996} established the persistent, viscous-dominated nature of individual vortex cores, but later research \citep{jimenez2020} revealed that large-scale organizations—specifically dipoles and streams—are the dominant carriers of kinetic energy. This progression marks a critical shift from viewing vortices as isolated building blocks to recognizing their collective dynamics as the primary governor of system energy. 

Despite the advances discussed above, research on the secondary instability of such coherent structures in two-dimensional turbulence remains limited. For three-dimensional shear flow, after the initial instability caused by Tollmien-Schlichting waves \citep{19}, the saturated nonlinear structures that develop become sensitive to three-dimensional perturbations, leading to secondary instability \citep{17,18} and ultimately direct transition to turbulence. Research on this nonlinear process and secondary instability generally relies on numerical simulations and continuation methods. For example, \cite{13}  used Keller's pseudo-arclength continuation method to compute nonlinear time-periodic equilibrium solutions in plane Poiseuille flow before chaos. Similarly, using spectral methods, \cite{orszag1980} investigated the development of finite-amplitude perturbations in three-dimensional channel flow, focusing on how three-dimensional perturbations induce transitions through nonlinear interactions. 
Notably, recent studies \citep{Chen2019, Chen2022} have demonstrated that large-scale vortical structures themselves can become unstable in high-speed flows, a finding that closely aligns with the motivation of the present work. Here, we adopt the classical linear stability eigenvalue approach to investigate the secondary instability of large-scale wavy structure in 2D channel.

In this work, direct numerical simulation (DNS) of the 2D Navier-Stokes equation is performed. 
To contrast the dynamics of stable and unstable regimes, two Reynolds number cases are studied, i.e. $Re$ = 3000 and $Re$ = 200000.
This choice is informed by the findings of \cite{chen2024}, who observed a transition in the streamwise velocity fluctuation scaling at $Re$ $\approx$ 10000; and by \cite{Falkovich2018}, who observed that below this $Re$ the flow field exhibits no turbulence characteristics, whereas above it turbulence are present. 
We employ DNS to evolve perturbations into saturated traveling wavy structures, which are subsequently extracted using singular value decomposition (SVD). Then, a linear stability analysis is developed on these extracted 2D structures to obtain their eigenvalue spectra, thereby characterizing their secondary instability. Note that this paper focuses on the stability analysis of large-scale structures rather than on how these structures are gradually formed (e.g., their nonlinear saturation process) in the simulation. Moreover, in contrast to classical secondary instability theory, which requires three-dimensional perturbations, the present study investigates the linear instability of the saturated two-dimensional traveling wave itself. It is demonstrated that this two-dimensional coherent structure can become linearly unstable at a sufficiently high $Re$, thus providing a direct pathway to turbulence at large Reynolds numbers. 

This paper is organized as follows. Section~\ref{Methods} describes the numerical methods and the SVD extraction process. Section~\ref{secondary instability} presents the stability analysis and eigenvalue spectra. Finally, Section~\ref{conclusion} summarizes the findings and suggests future research directions.

\begin{figure}
    \centering
    \includegraphics[width=0.6\textwidth]{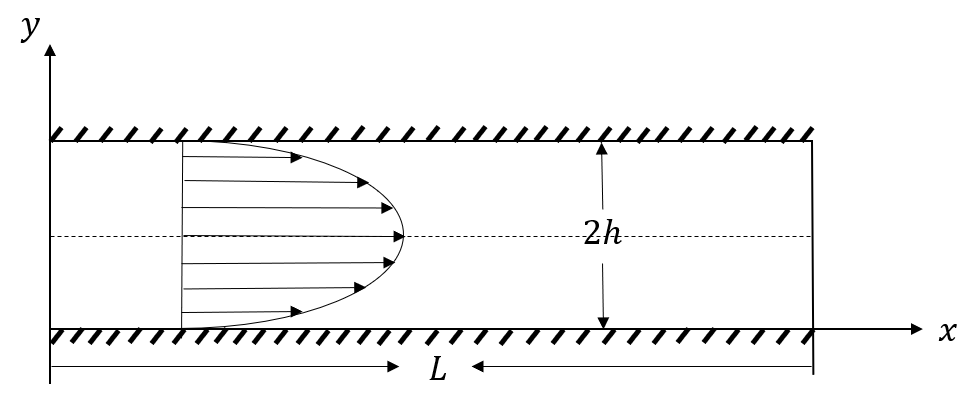}
    \caption{Sketch of the computational domain. Here, $h$ is the half height of the channel, and $L$ is the length in the streamwise direction.}
    \label{fig:60}
\end{figure}

\section{Methods}
\label{Methods}
\subsection{Simulations}
The 2D incompressible flow between two parallel plates driven by a pressure difference is considered. 
The computational domain is shown in figure~\ref{fig:60}. 
The perturbation equation is:
\begin{equation}
\partial_t \mathbf{u} + (\mathbf{U_0} + \mathbf{u}) \cdot \nabla (\mathbf{U_0} + \mathbf{u}) = -\nabla p + \frac{1}{Re} \nabla^2 \mathbf{u},
\label{1}
\end{equation}
\begin{equation}
\nabla \cdot \mathbf{u} = 0.
\end{equation}
Here, the bulk Reynolds number $Re$ is defined as \( Re = U_m h/\nu \), where \( \nu \) is the fluid kinematic viscosity, and $U_m$ is the bulk velocity. 
The dimensionless coordinates of the top and bottom walls are respectively \( y = \pm 1 \), where the non-slip boundary conditions are applied. 
The dimensionless basic flow profile is \( \mathbf{U_0} = (U, V) = (1.5 - y^2, 0) \) and \( \mathbf{u} = (u, v) \) is the perturbation velocity. Hereafter, all the quantities are non-dimensionalized.

\begin{figure}
    \centering
    \begin{subfigure}{0.7\textwidth}
        \centering
        \begin{overpic}[width=\linewidth, trim=-10pt 0 -10pt 0, clip]{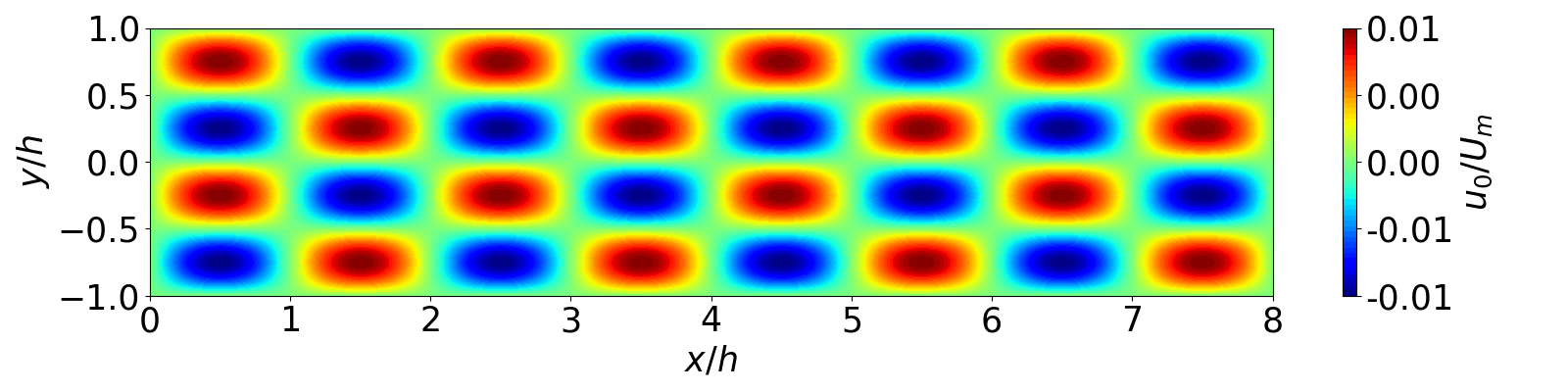}
            \put(-5,20){\small(a)}
        \end{overpic}
    \end{subfigure}
    \begin{subfigure}{0.7\textwidth}
        \centering
        \begin{overpic}[width=\linewidth, trim=-10pt 0 -10pt 0, clip]{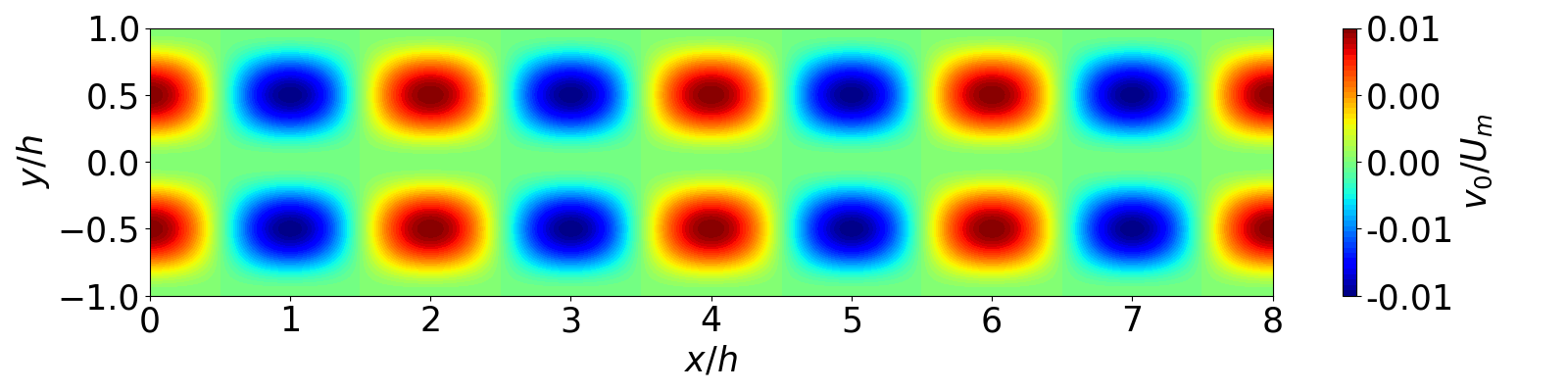}
            \put(-5,20){\small(b)}
        \end{overpic}
    \end{subfigure}\\
    \caption{Contours of initial perturbation velocity \( \mathbf{u_0} = (u_0, v_0)\). (a) The streamwise initial perturbation velocity $u_0/U_m$. (b) The initial perturbation velocity in the wall-normal direction $v_0/U_m$.}
    \label{uv_0}
\end{figure}

DNS of the 2D Navier-Stokes equations is performed using our finite-differential code \citep{xie2021low}. The details of our DNS settings are described in the Appendix~\ref{Numerical Setup}. 
Adding the perturbations at the initial moment, the large-scale traveling wavy structures formed after the sufficiently developed perturbations. 
Following \cite{Falkovich2018}, the initial perturbation velocity \( \mathbf{u_0} = (u_0, v_0)\) is given by:

\begin{equation}
    \mathit{u}_0 = -2\left(\frac{a}{k_x}\right)\sin(k_y y)\cos(k_y y)\sin(k_x x)
    \label{m}
    \end{equation}
    \begin{equation}
    \mathit{v}_0 = \left(\frac{a}{k_y}\right)\sin(k_y y)\sin(k_y y)\cos(k_x x)
    \label{n}
\end{equation}
where \( k_x = \pi n/L \), \( k_y = \pi m/L \), with \( n,m = 1,2,3,\ldots \). 
Note that the perturbations vanish at the boundaries, satisfying the non-slip boundary conditions. 
Here, similar to \cite{Falkovich2018}, we set \( m = n = 8 \) and  \(a/k_x = a/k_y = 0.1\). The corresponding flow field contour is shown in figure~\ref{uv_0}. 
We note that \cite{chen2024} conducted simulations with initial random noise perturbations. After the flow became fully developed, they also observed travelling wavy structures, indicating that the formation of large-scale structures ultimately depends only on the Reynolds number and not on the specific form of the initial perturbations.

Figure~\ref{RE_111111} shows the variation of the skin friction coefficient $C_f$ with $Re$. The present DNS data show excellent agreement with the theoretical scaling $C_f \sim Re^{-1/2}$ and existing data \citep{Falkovich2018, kellay2012testing, tran2010macroscopic}, confirming the reliability of the numerical simulations and providing a trustworthy data basis for the subsequent flow field analysis.

\begin{figure}
    \centering
    \hspace*{-1.2cm}
    \includegraphics[width=0.7\textwidth]{data_CF.png}
    \caption{$C_f$ varying with $Re$ in 2D channel flows. The solid line represents the 2D friction law $Re$\(^{-1/2}\). Black squares represent DNS data from \cite{Falkovich2018}, yellow crosses are experimental data from \cite{kellay2012testing} and brown crosses are from \cite{tran2010macroscopic}. The two red triangles indicate the present DNS results obtained at $Re$ = 3000 and $Re$ = 200000.}
    \label{RE_111111}
\end{figure}

\begin{figure}
    \centering
    \begin{subfigure}{0.45\textwidth}
        \centering
        \begin{overpic}[width=\linewidth, trim=-10pt 0 -10pt 0, clip]{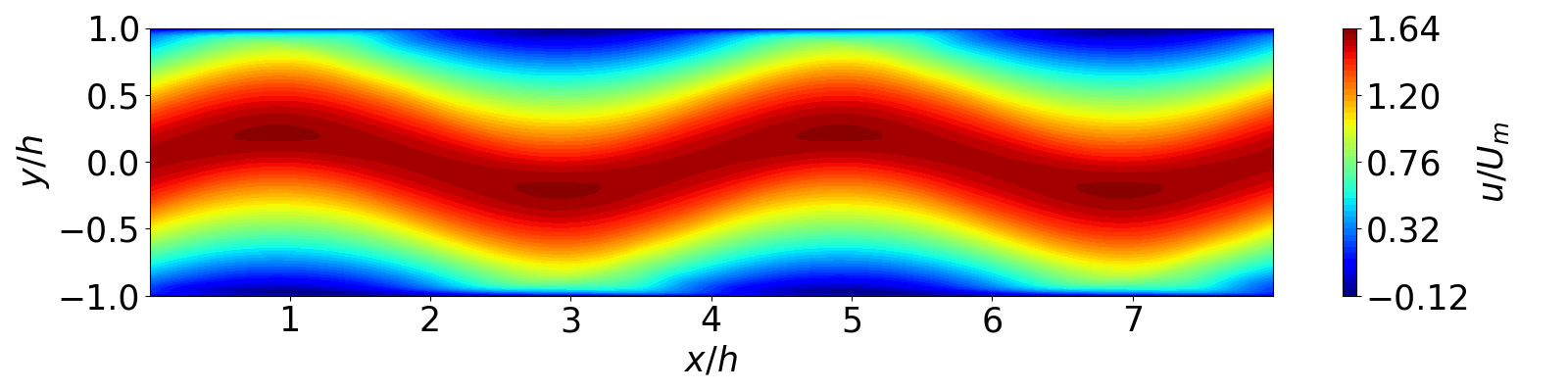}
            \put(-5,20){\small(a)}
        \end{overpic}
    \end{subfigure}
    \begin{subfigure}{0.45\textwidth}
        \centering
        \begin{overpic}[width=\linewidth, trim=-10pt 0 -10pt 0, clip]{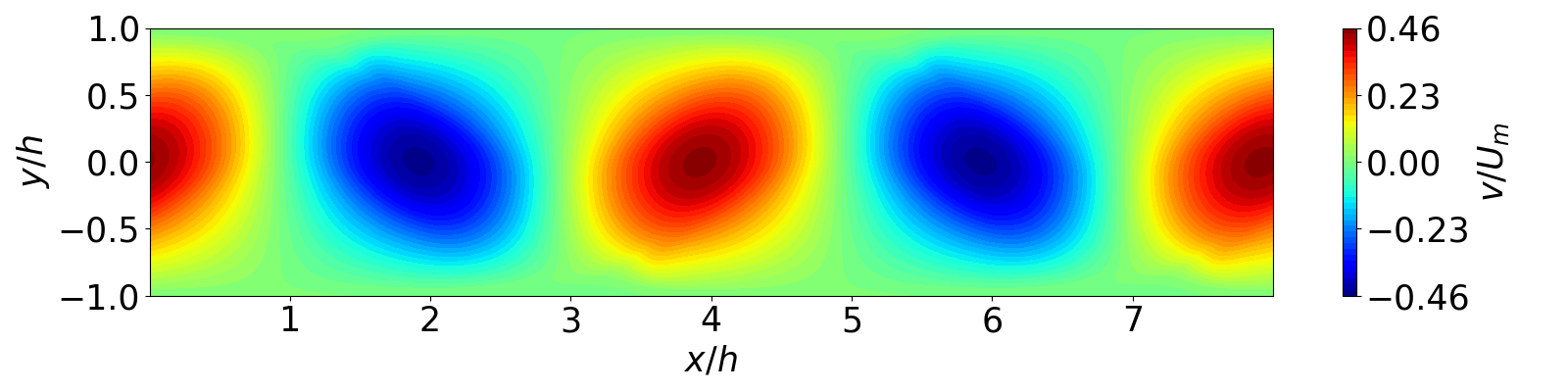}
            \put(-5,20){\small(b)}
        \end{overpic}
    \end{subfigure}\\
    \begin{subfigure}{0.45\textwidth}
        \centering
        \begin{overpic}[width=\linewidth, trim=-10pt 0 -10pt 0, clip]{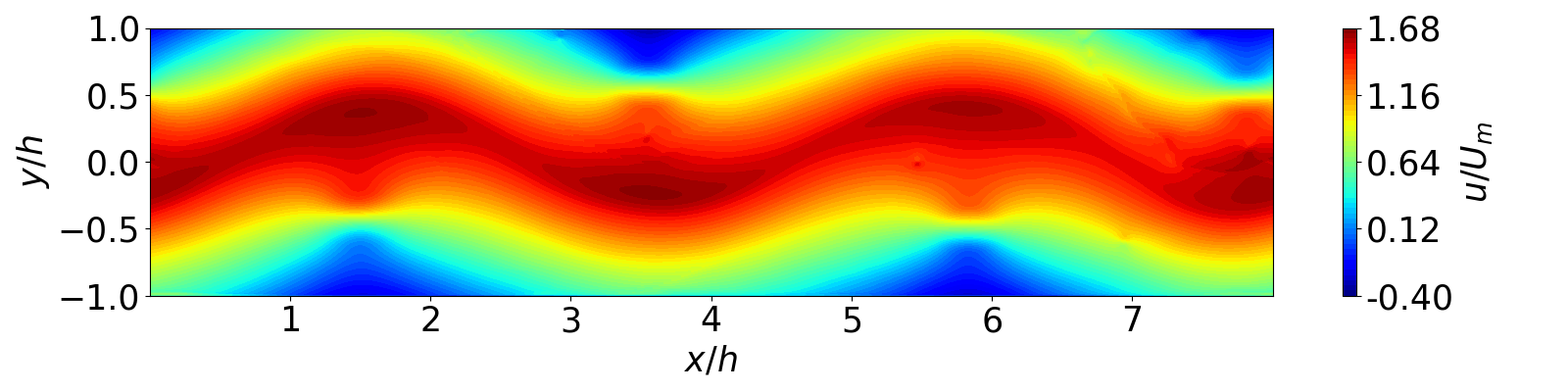}
            \put(-5,20){\small(c)}
        \end{overpic}
    \end{subfigure}
    \begin{subfigure}{0.45\textwidth}
        \centering
        \begin{overpic}[width=\linewidth, trim=-10pt 0 -10pt 0, clip]{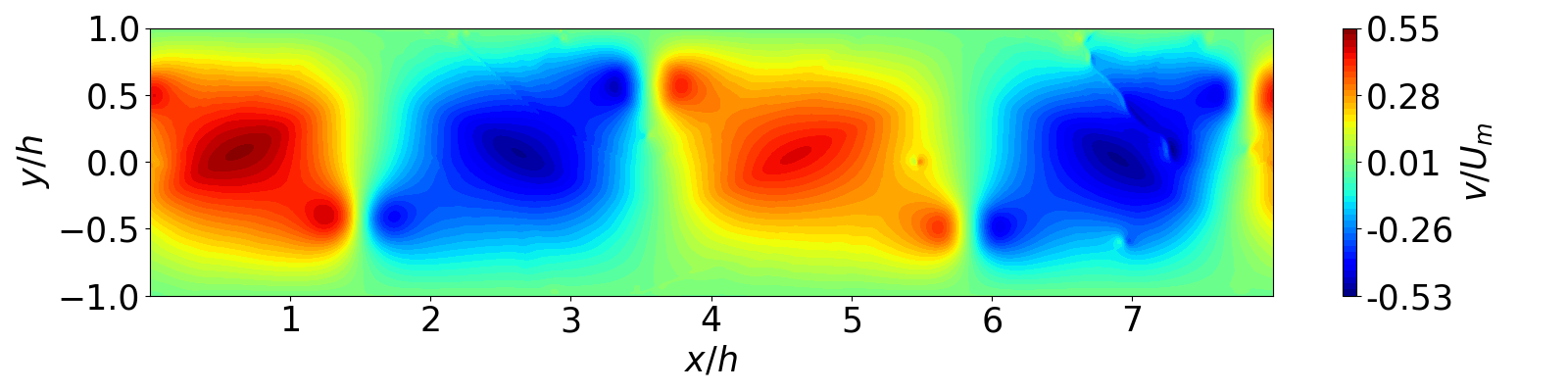}
            \put(-5,20){\small(d)}
        \end{overpic}
    \end{subfigure}
    \caption{Velocity contours of fully developed 2D channel flows. (a) The dimensionless streamwise velocity \( u/U_m \) at $Re$ = 3000. (b) The dimensionless wall-normal velocity \( v/U_m \) at $Re$ = 3000. (c) \( u/U_m \) at $Re$ = 200000. (d) \( v/U_m \) at $Re$ = 200000.
    }
    \label{s}
\end{figure}

The fully developed flow fields are shown in figure~\ref{s}.
At both Reynolds numbers, it is clearly observed that large-scale traveling wavy structures exist. 
The difference is that at the lower Reynolds number, $Re$ = 3000, the flow field does not appear to exhibit turbulent characteristics (small scale velocity fluctuations with low energy proportion) (figure~\ref{s}(a, b)), indicating that the large-scale traveling wavy structure may be stable.
However, at $Re$ = 200000, the flow field clearly exhibits turbulent characteristics (figure~\ref{s}c,~d), indicating that the large-scale traveling wavy structure may have become unstable. 
In order to obtain a pure large-scale traveling wavy structure in the flow field for subsequent stability analysis, the SVD method needs to be used for removing the small scale velocity fluctuations with low energy proportion.

\subsection{Singular value decomposition}
The singular value decomposition \citep{SVD} is employed herein as an optimal modal decomposition technique to filter high-wavenumber velocity fluctuations and isolate the coherent, large-scale traveling wavy structures from the DNS data.
For a discrete velocity field represented by a matrix $A \in \mathbb{R}^{m \times n}$, the SVD is defined as:
\begin{equation}
A = P \Sigma Q^T
\end{equation}
where $P \in \mathbb{R}^{m \times m}$ and $Q \in \mathbb{R}^{n \times n}$ are orthogonal matrices containing the left and right singular vectors, respectively, and $\Sigma \in \mathbb{R}^{m \times n}$ is a diagonal matrix of singular values $\sigma_i$ ordered such that $\sigma_1 \geq \sigma_2 \geq \dots \geq \sigma_r > 0$.
The detailed derivations for the SVD method can be found in Appendix~\ref{Complete Derivation of SVD}.

In this paper, matrix \( A \) is used to store the data of velocity \( u \) or \( v \).
To extract the dominant physical features, we employ a rank-$k$ approximation, $\hat{A}$, by  truncating only the top \( k \) largest singular values and their corresponding singular vectors:

\begin{equation}
\hat{A}=P_k \Sigma_k Q_k^T=\sum_{i=1}^k \sigma_i\mathbf{p_i}\mathbf{q_i^T},
\end{equation}
where $\hat{A}\approx A$ for rank-\( k \) approximation, \( P_k \in \mathbb{R}^{m \times k} \) is the first \( k \) left singular vectors $\mathbf{p_i}$, \( \Sigma_k \in \mathbb{R}^{k \times k} \) is the diagonal matrix composed of the first \( k \) singular values $\sigma_i$, \( Q_k \in \mathbb{R}^{n \times k} \) is the first \( k \) right singular vectors $\mathbf{q_i}$.

\begin{figure}
    \centering
    \begin{subfigure}{0.45\textwidth}
        \centering
        \begin{overpic}[width=\linewidth, trim=-10pt 0 -10pt 0, clip]{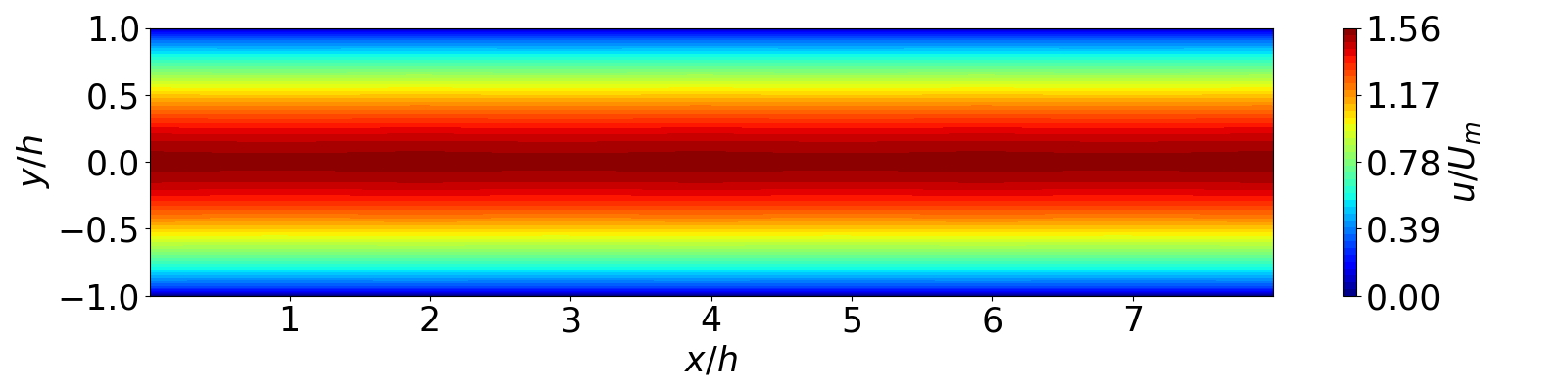}
            \put(-5,20){\small(a)}
        \end{overpic}
    \end{subfigure}
    \begin{subfigure}{0.45\textwidth}
        \centering
        \begin{overpic}[width=\linewidth, trim=-10pt 0 -10pt 0, clip]{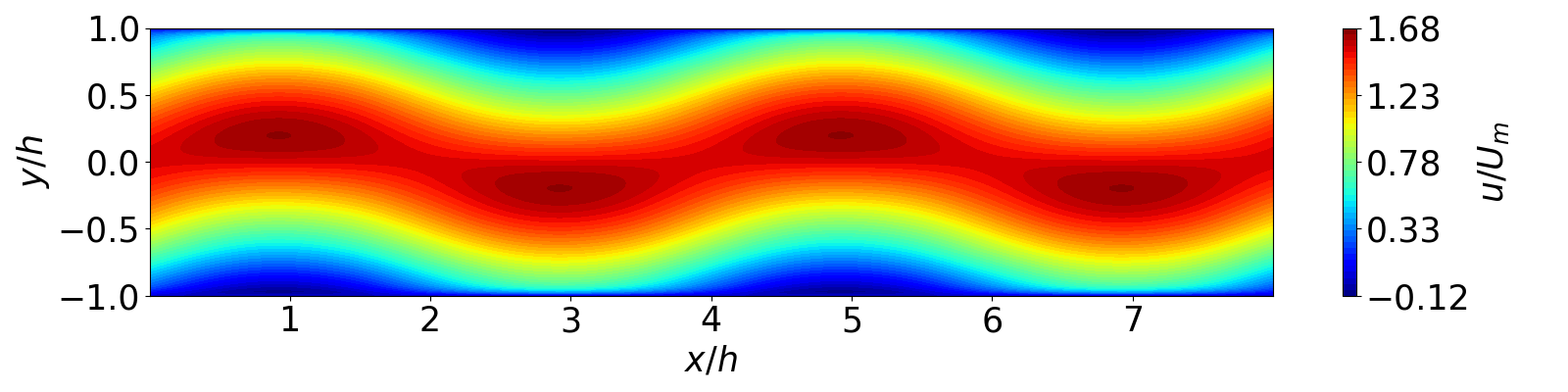}
            \put(-5,20){\small(b)}
        \end{overpic}
    \end{subfigure}\\
    \begin{subfigure}{0.45\textwidth}
        \centering
        \begin{overpic}[width=\linewidth, trim=-10pt 0 -10pt 0, clip]{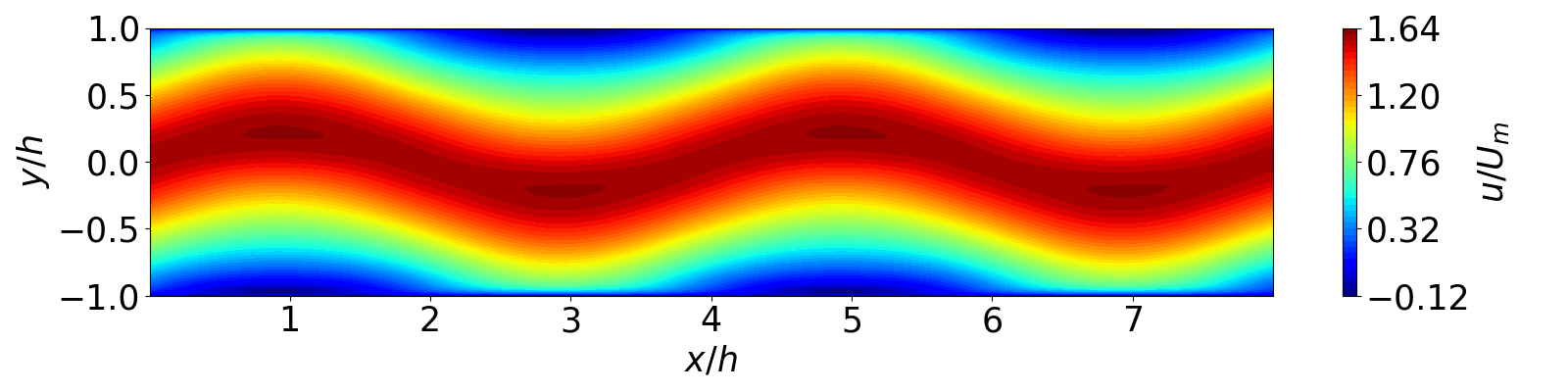}
            \put(-5,20){\small(c)}
        \end{overpic}
    \end{subfigure}
    \begin{subfigure}{0.45\textwidth}
        \centering
        \begin{overpic}[width=\linewidth, trim=-10pt 0 -10pt 0, clip]{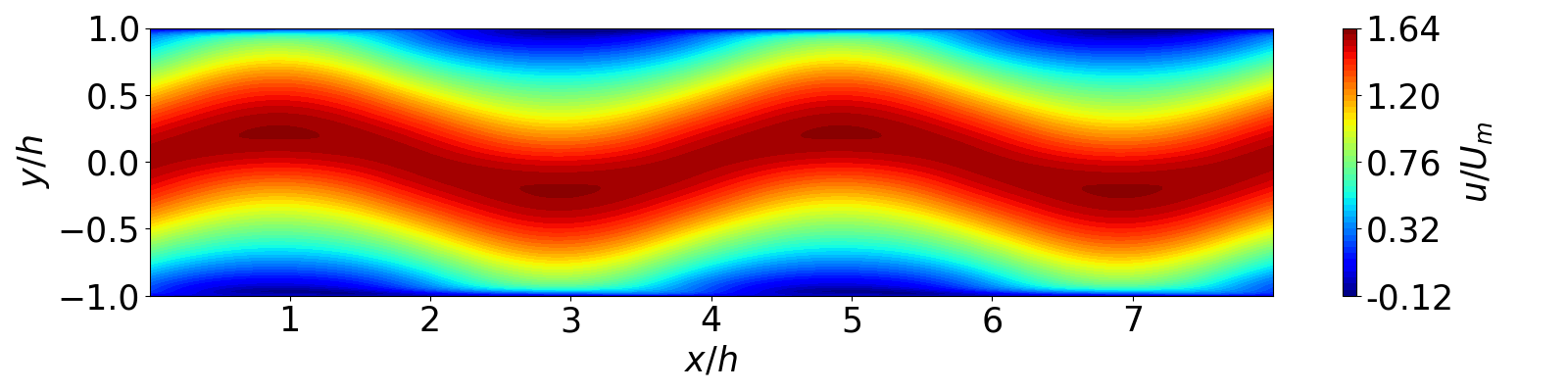}
            \put(-5,20){\small(d)}
        \end{overpic}
    \end{subfigure}
    \caption{First four rank SVD approximations on its dimensionless streamwise velocity \( u/U_m \) at $Re$ = 3000 (a): rank-1 approximation (b): rank-2 approximation (c): rank-3 approximation (d): rank-4 approximation.}
    \label{make}
\end{figure}

\begin{figure}
    \centering
    \begin{subfigure}{0.45\textwidth}
        \centering
        \begin{overpic}[width=\linewidth, trim=-10pt 0 -10pt 0, clip]{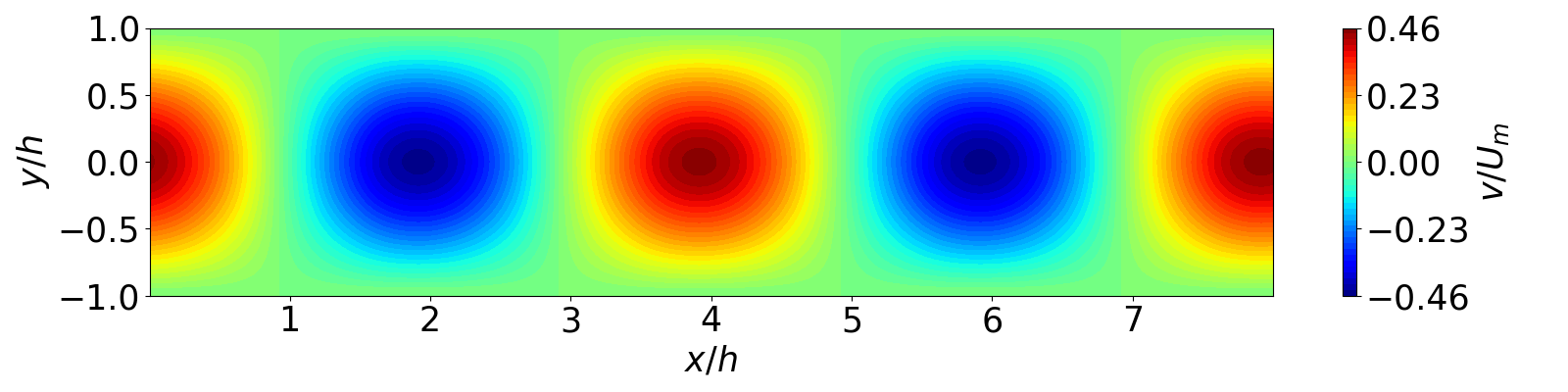}
            \put(-5,20){\small(a)}
        \end{overpic}
    \end{subfigure}
    \begin{subfigure}{0.45\textwidth}
        \centering
        \begin{overpic}[width=\linewidth, trim=-10pt 0 -10pt 0, clip]{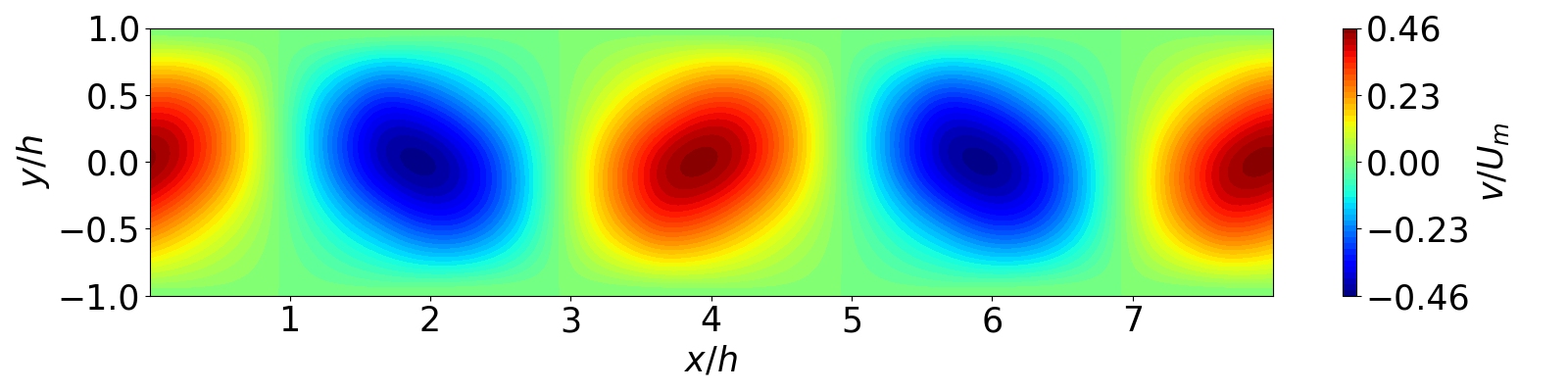}
            \put(-5,20){\small(b)}
        \end{overpic}
    \end{subfigure}\\
    \begin{subfigure}{0.45\textwidth}
        \centering
        \begin{overpic}[width=\linewidth, trim=-10pt 0 -10pt 0, clip]{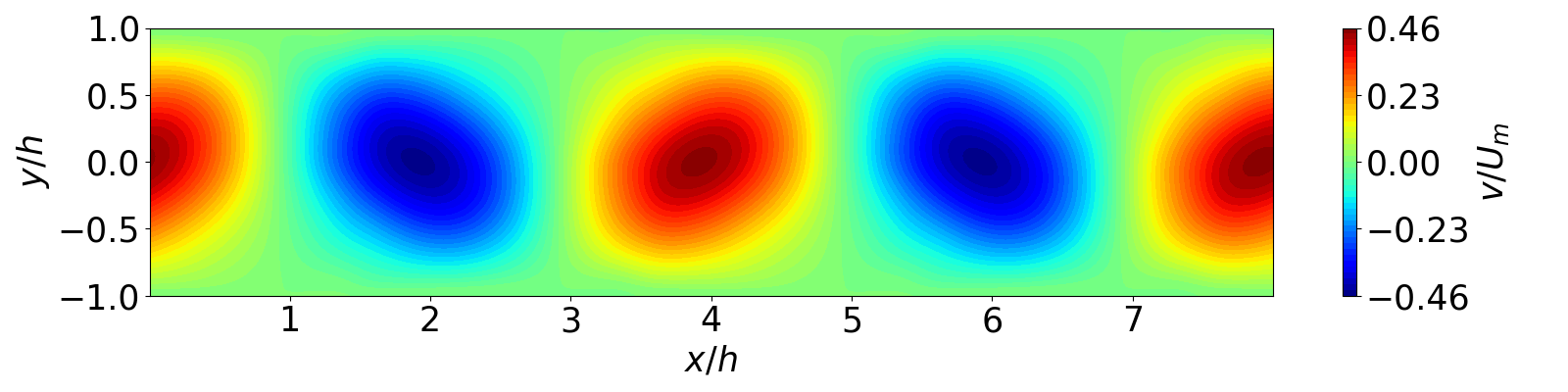}
            \put(-5,20){\small(c)}
        \end{overpic}
    \end{subfigure}
    \begin{subfigure}{0.45\textwidth}
        \centering
        \begin{overpic}[width=\linewidth, trim=-10pt 0 -10pt 0, clip]{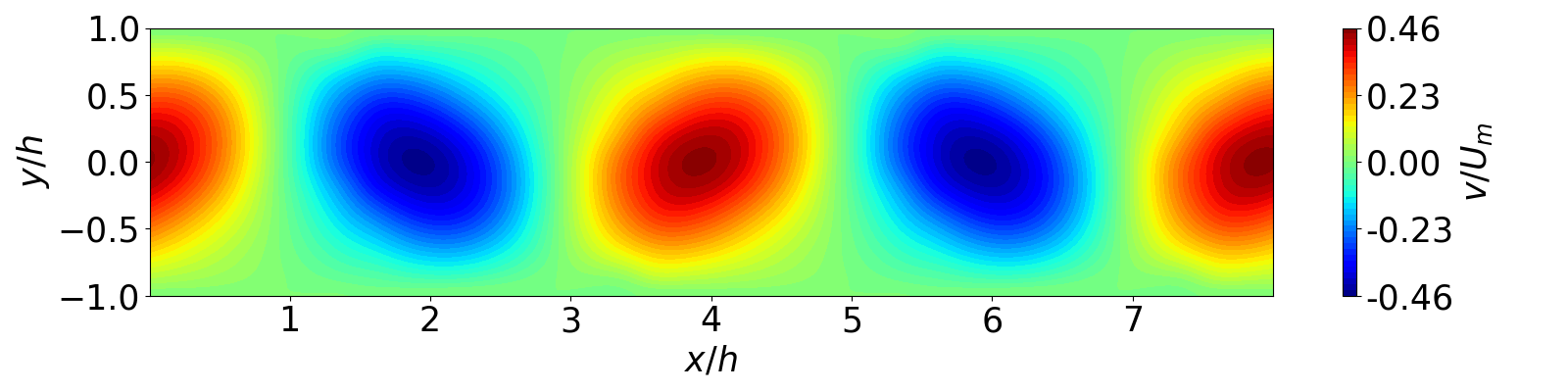}
            \put(-5,20){\small(d)}
        \end{overpic}
    \end{subfigure}
    \caption{First four rank SVD approximations on its dimensionless streamwise velocity \( v/U_m \) at $Re$ = 3000 (a): rank-1 approximation (b): rank-2 approximation (c): rank-3 approximation (d): rank-4 approximation.}
    \label{gongsunli}
\end{figure}

At $Re = 3000$, the first four rank SVD approximations of the dimensionless streamwise ($u/U_m$) and wall-normal ($v/U_m$) velocities (figures \ref{make} and \ref{gongsunli}) exhibit negligible differences from the original instantaneous snapshots (figure~\ref{s}a). 
Quantitative analysis of the cumulative energy proportion (table \ref{energy proportions at $Re$ = 3000}) reveals that the primary modes capture nearly 99\% of the total energy. 
This suggests that at $Re = 3000$, the flow lacks the multi-scale characteristics of turbulence and remains essentially laminar.
Consequently, for the subsequent stability analysis at this Reynolds number, the unfiltered DNS data is used directly to represent the base flow.

\begin{table}
  \begin{center}
\def~{\hphantom{0}}
  \begin{tabular}{lcccc}
                   & rank-1  & rank-2  & rank-3  & rank-4  \\[3pt]
       \( u/U_m \) & 91.52\% & 99.03\% & 99.89\% & 99.99\% \\
       \( v/U_m \) & 98.71\% & 99.64\% & 99.98\% & 99.99\% \\
  \end{tabular}
  \caption{Energy proportion under the first four rank SVD approximations at $Re$ = 3000. The energy proportion under the rank-\(k\) approximation is \((\sum_{i=1}^k \sigma_i) / (\sum_{i=1}^r \sigma_i)\).}
  \label{energy proportions at $Re$ = 3000}
  \end{center}
\end{table}

\begin{figure}
    \centering
    \begin{subfigure}{0.45\textwidth}
        \centering
        \begin{overpic}[width=\linewidth, trim=-10pt 0 -10pt 0, clip]{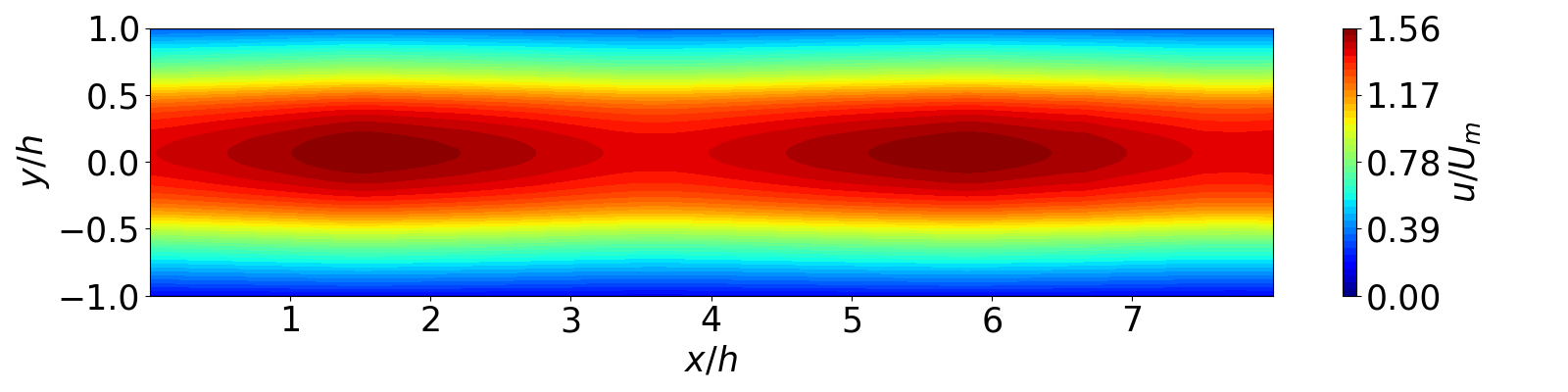}
            \put(-5,20){\small(a)}
        \end{overpic}
    \end{subfigure}
    \begin{subfigure}{0.45\textwidth}
        \centering
        \begin{overpic}[width=\linewidth, trim=-10pt 0 -10pt 0, clip]{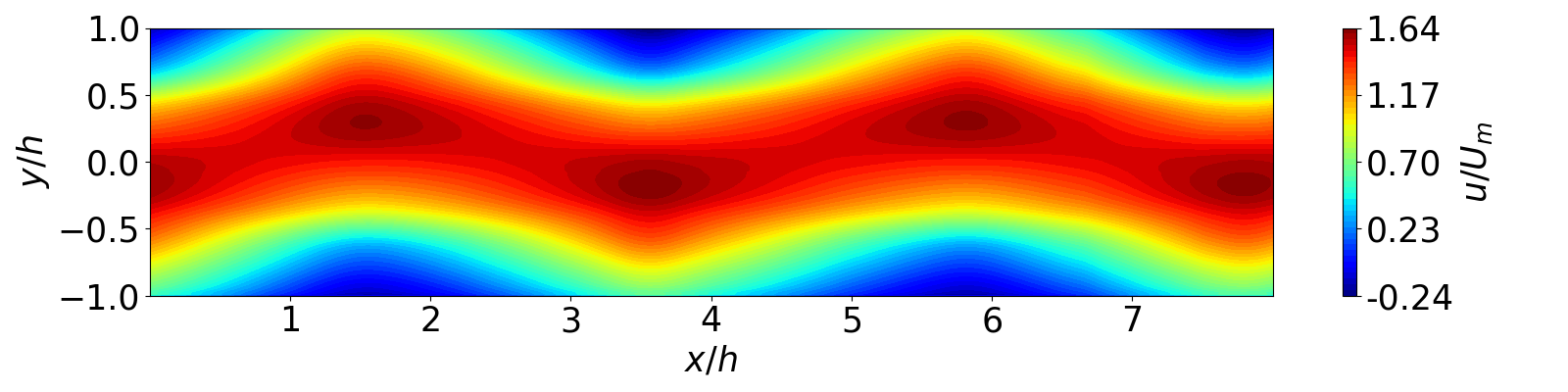}
            \put(-5,20){\small(b)}
        \end{overpic}
    \end{subfigure}\\
    \begin{subfigure}{0.45\textwidth}
        \centering
        \begin{overpic}[width=\linewidth, trim=-10pt 0 -10pt 0, clip]{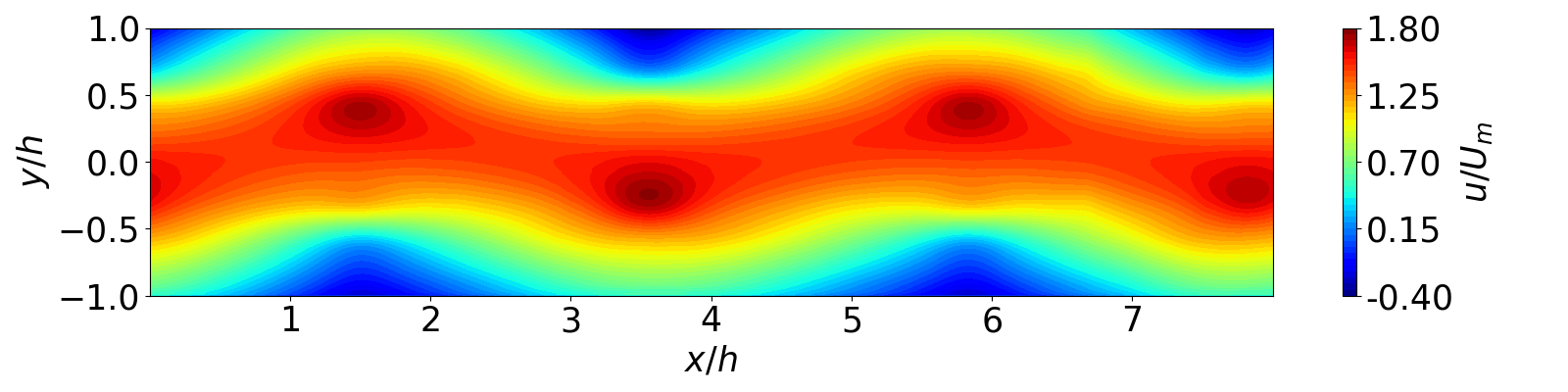}
            \put(-5,20){\small(c)}
        \end{overpic}
    \end{subfigure}
    \begin{subfigure}{0.45\textwidth}
        \centering
        \begin{overpic}[width=\linewidth, trim=-10pt 0 -10pt 0, clip]{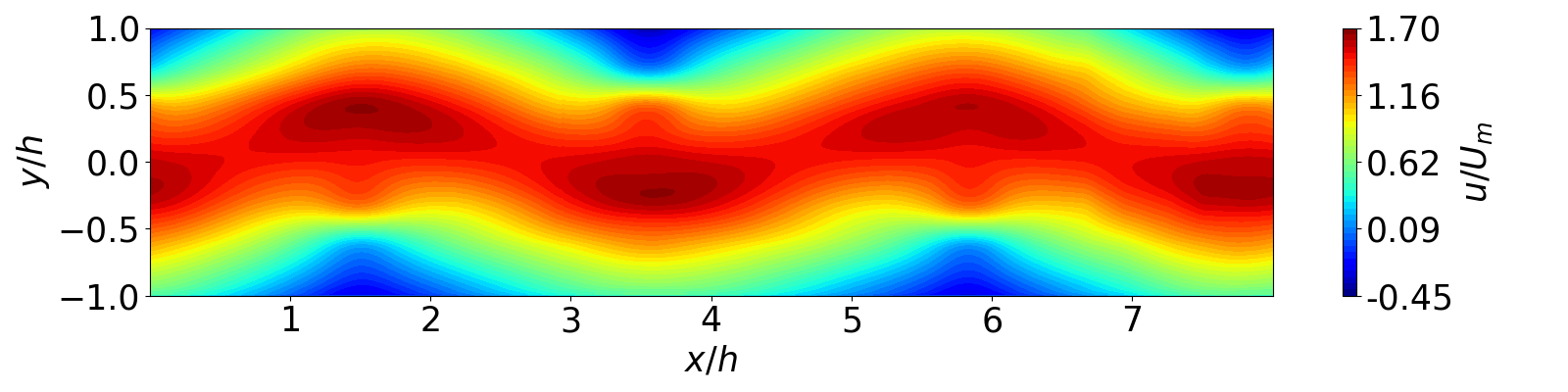}
            \put(-5,20){\small(d)}
        \end{overpic}
    \end{subfigure}
    \caption{First four rank SVD approximations on its dimensionless streamwise velocity \( u/U_m \) at $Re$ = 200000 (a): rank-1 approximation (b): rank-2 approximation (c): rank-3 approximation (d): rank-4 approximation}
    \label{lan}
\end{figure}

\begin{figure}
    \centering
    \begin{subfigure}{0.45\textwidth}
        \centering
        \begin{overpic}[width=\linewidth, trim=-10pt 0 -10pt 0, clip]{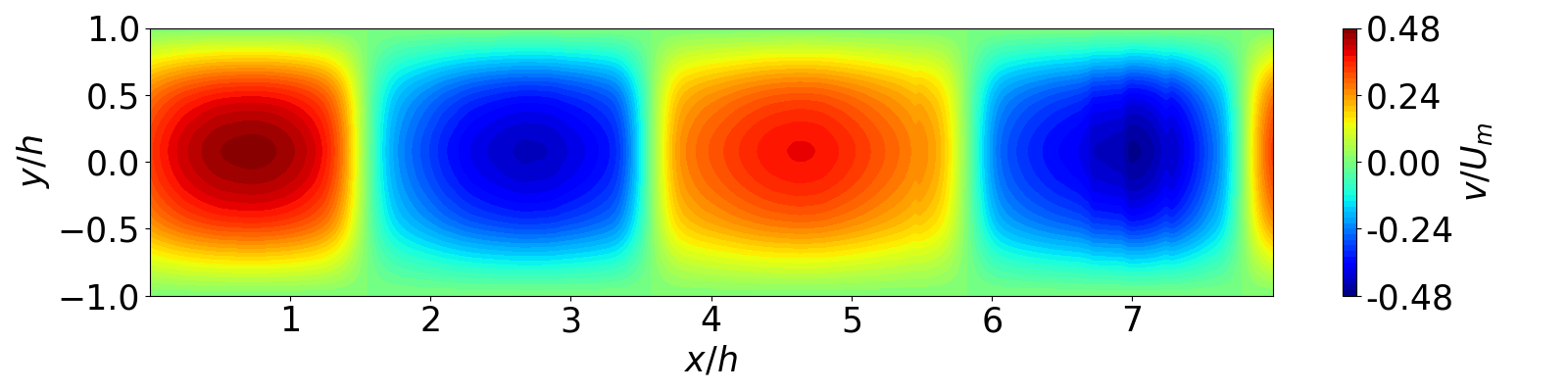}
            \put(-5,20){\small(a)}
        \end{overpic}
    \end{subfigure}
    \begin{subfigure}{0.45\textwidth}
        \centering
        \begin{overpic}[width=\linewidth, trim=-10pt 0 -10pt 0, clip]{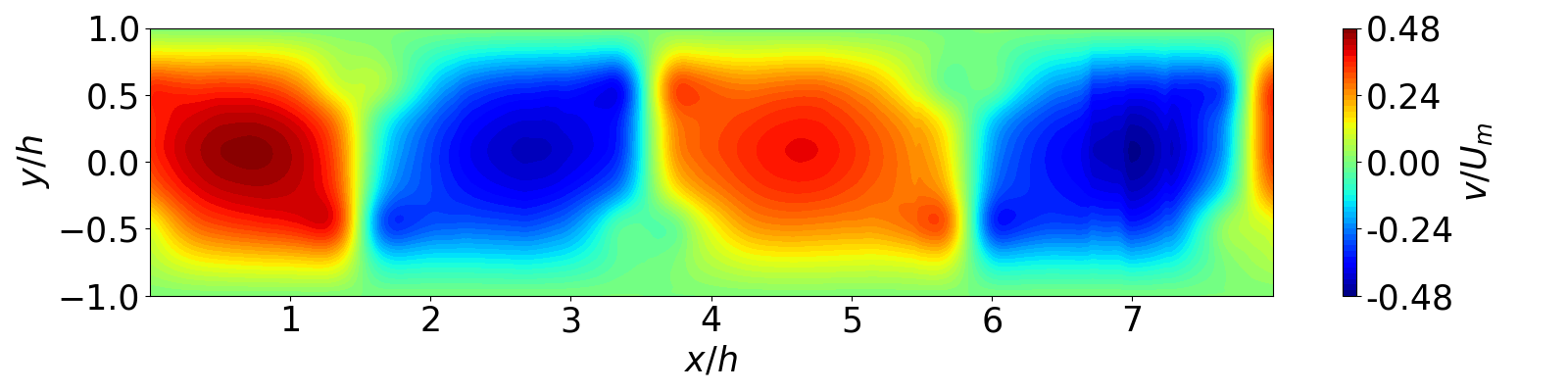}
            \put(-5,20){\small(b)}
        \end{overpic}
    \end{subfigure}\\
    \begin{subfigure}{0.45\textwidth}
        \centering
        \begin{overpic}[width=\linewidth, trim=-10pt 0 -10pt 0, clip]{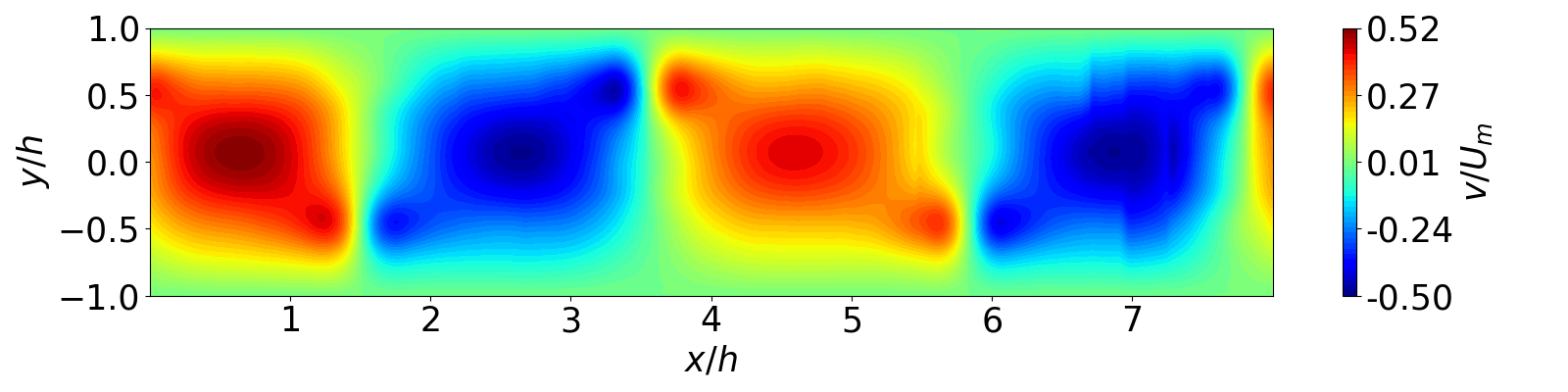}
            \put(-5,20){\small(c)}
        \end{overpic}
    \end{subfigure}
    \begin{subfigure}{0.45\textwidth}
        \centering
        \begin{overpic}[width=\linewidth, trim=-10pt 0 -10pt 0, clip]{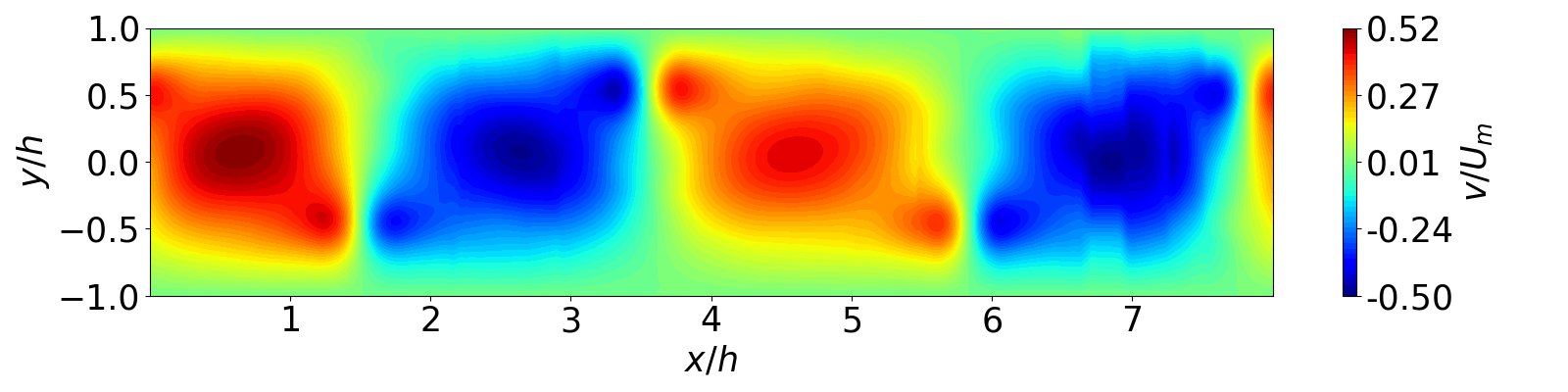}
            \put(-5,20){\small(d)}
        \end{overpic}
    \end{subfigure}
    \caption{First four rank SVD approximations on its dimensionless normal velocity \( v/U_m \) at $Re$ = 200000 (a): rank-1 approximation (b): rank-2 approximation (c): rank-3 approximation (d): rank-4 approximation}.
    \label{yao}
\end{figure}

In contrast, at $Re$ = 200000, the SVD approximations of the instantaneous flow field exhibit a broad range of scales (figures \ref{lan} and \ref{yao}). The energy proportions under the first four rank SVD approximations (table \ref{energy proportions at $Re$ = 200000}) indicate that the first two modes account for the vast majority of the kinetic energy, while higher-order modes represent low-energy velocity fluctuations and numerical noise.

\begin{table}
  \begin{center}
\def~{\hphantom{0}}
  \begin{tabular}{lcccc}
                   & rank-1  & rank-2  & rank-3  & rank-4  \\[3pt]
       \( u/U_m \) & 80.12\% & 96.07\% & 98.82\% & 99.77\% \\
       \( v/U_m \) & 94.23\% & 97.86\% & 98.97\% & 99.91\% \\
  \end{tabular}
  \caption{Energy proportion under the first four rank SVD approximations at $Re$=200000. The energy proportion under the rank-\( k \) approximation is $(\sum_{i=1}^k \sigma_i)/(\sum_{i=1}^r \sigma_i)$.}
  \label{energy proportions at $Re$ = 200000}
  \end{center}
\end{table}

\begin{figure}
    \centering
    \begin{subfigure}{0.45\textwidth}
        \centering
        \begin{overpic}[width=\linewidth, trim=-10pt 0 -10pt 0, clip]{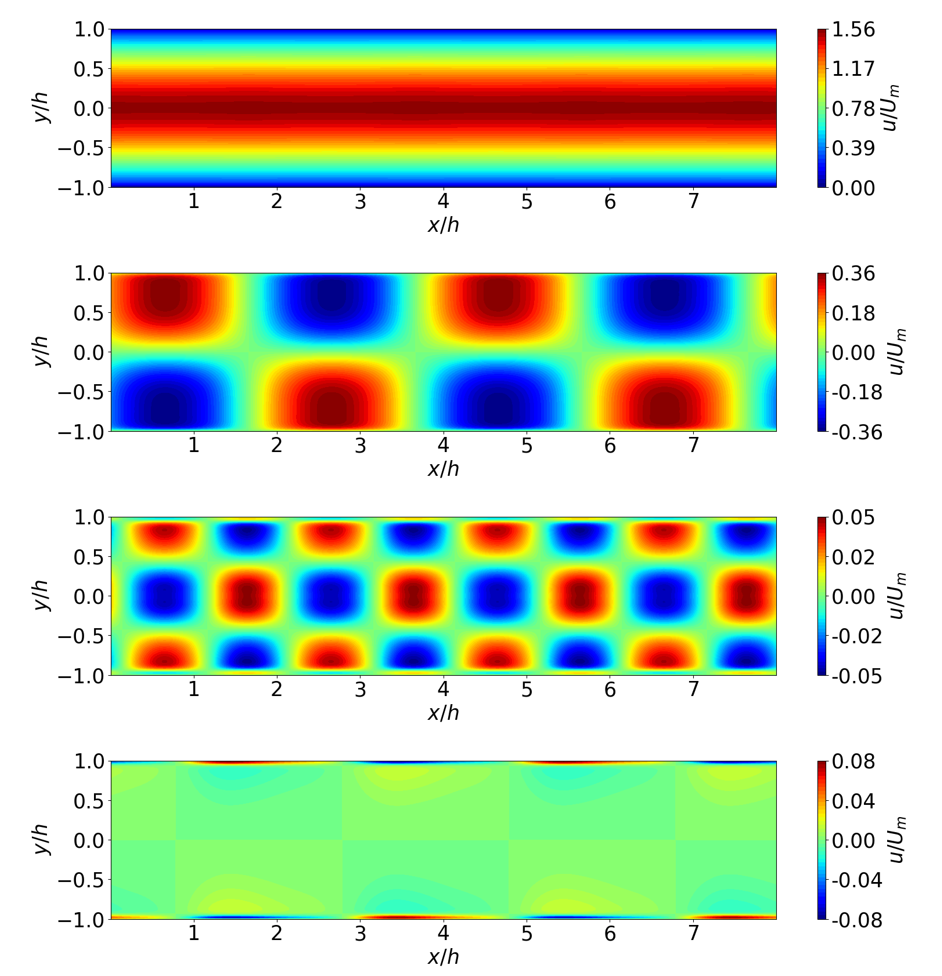}
            \put(-5,92){\small(a)}
        \end{overpic}
    \end{subfigure}
    \begin{subfigure}{0.45\textwidth}
        \centering
        \begin{overpic}[width=\linewidth, trim=-10pt 0 -10pt 0, clip]{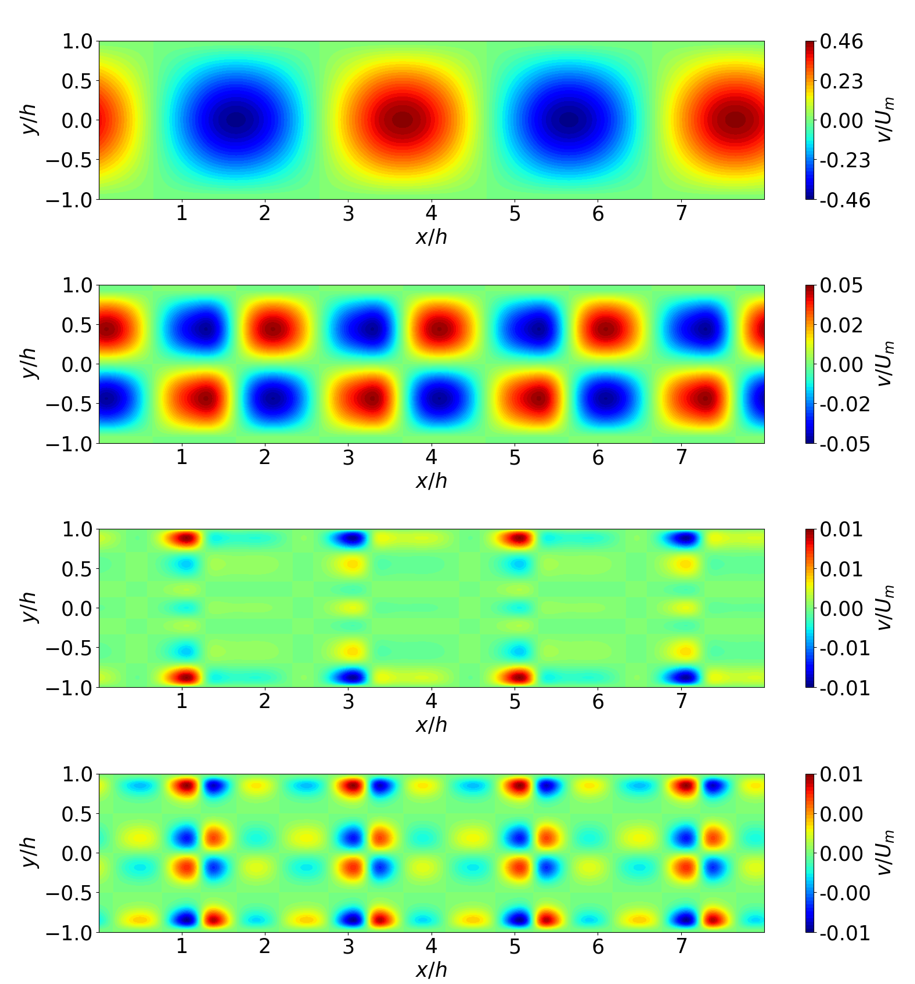}
            \put(-5,90){\small(b)}
        \end{overpic}
    \end{subfigure}
    \caption{Each of the first four SVD modes of $u$ and $v$ at $Re=3000$. (a) From top to bottom are the first, second, third, and fourth SVD modes of the streamwise velocity $u/U_m$, respectively. (b) From top to bottom are the first, second, third, and fourth SVD modes of the normal velocity $v/U_m$, respectively.}
    \label{gejie}
\end{figure}

The rank-2 approximation already captures the vast majority of the energy. However, for the wall normal velocity \( v/U_m \), the rank-1 approximation already contains the vast majority of energy. This is because, in the SVD, $\sigma_1$ usually corresponds to the background average flow of the flow field, while $\sigma_2$ corresponds to a certain flow field structure. Taking $Re$ = 3000 as an example, each of the first four SVD modes is presented in figure~\ref{gejie}. It can be observed that for the velocity $u$, the first SVD mode essentially corresponds to the initial mean flow velocity $U$ with a parabolic profile, while the second mode represents the most dominant perturbation structure, and the higher-order modes correspond to higher-order perturbation structures. For the velocity $v$, since the initial mean normal velocity $V$ is zero, the first SVD mode directly represents the most dominant perturbation structure. 
Furthermore, these figures also illustrate that the formation of large-scale structures results from complex nonlinear interactions between the initial perturbations and the mean flow, rather than simply from the amplification of the initial perturbation structures (figure~\ref{uv_0}). In the present analysis, it is this large-scale traveling wavy structure, but the average flow does not contain normal velocity. Therefore, $\sigma_1$ of the normal velocity \( v/U_m \) corresponds to this large-scale traveling wavy structure. 
Based on this energy partitioning, we select a rank-2 approximation for $u/U_m$ and a rank-1 approximation for $v/U_m$ to reconstruct the coherent base state for the secondary instability analysis.

\section{Secondary instability}
\label{secondary instability}
Although the present investigation is restricted to a two-dimensional configuration, it is important to emphasize that secondary instability theory provides a robust mathematical framework for analyzing the linear stability of unsteady or spatially-periodic velocity profiles, such as the traveling wavy structures identified herein. In this 2D context, the base flow is characterized by a time-dependent (or phase-dependent) velocity profile. 
Specifically, the previously steady base flow profile now becomes unsteady, taking the form:

\begin{equation}
\mathbf{U}_0 = (U, V)+(\widetilde{u^{\mathrm{2D}}}, \widetilde{v^{\mathrm{2D}}}),
\label{U_profile}
\end{equation}
where $\widetilde{u^{\mathrm{2D}}}$ and $\widetilde{v^{\mathrm{2D}}}$ represent the streamwise velocity and normal velocity at each point on large-scale traveling wavy structures, respectively. For convenience, a Galilean transformation \(x' = x - Ct\) anchors the coordinate system to the large-scale traveling wavy structure in order to analyze the problem under a steady background velocity, as in classical linear stability analysis:

\begin{equation}
    \mathbf{U}_0^{\prime}=[U-C+\widetilde{u^{\mathrm{2D}}}(x', y),\widetilde{v^{\mathrm{2D}}}(x', y)],
    \label{New_U_profile}
\end{equation}
\( C \) is the wave speed, obtained by dividing the frequency $\omega$ by the streamwise wave number $\alpha$, and where: 

\begin{equation}
\widetilde{u^{\mathrm{2D}}}=u^{\mathrm{2D}}(y)e^{i\alpha x^\prime}+u^{\mathrm{2D}}(y)e^{-i\alpha x^\prime}
\end{equation}

\begin{equation}
\widetilde{v^{\mathrm{2D}}}=v^{\mathrm{2D}}(y)e^{i\alpha x^\prime}+v^{\mathrm{2D}}(y)e^{-i\alpha x^\prime}
\end{equation}
$u^{\mathrm{2D}}$ and $v^{\mathrm{2D}}$ are the amplitudes of the $\widetilde{u^{\mathrm{2D}}}$ and $\widetilde{v^{\mathrm{2D}}}$, respectively, and are positive real quantities. One hundred instantaneous flow field snapshots after SVD approximation were selected  for Fourier transform to obtain the nondimensional streamwise wave numbers $\hat{\alpha}$ and frequencies $\hat{\omega}$ of the large-scale traveling wavy structures, as shown in Table~\ref{66677}. 

\begin{table}
  \begin{center}
\def~{\hphantom{0}}
  \begin{tabular}{lcccc}
    $Re$ &  $Re_\tau$ & $\hat{\alpha} = \alpha h$ & $\hat{\omega} = \omega h/U_m$ & $\hat{C}=\omega/\alpha U_m$   \\[3pt]
       3000 & 127  & 1.570 & 1.256 & 0.803   \\
     200000 & 2841 & 1.461 & 0.019 & 0.013  \\
  \end{tabular}
  \caption{Dimensionless streamwise wave number $\hat{\alpha}$, frequency $\hat{\omega}$, wave speed $\hat{C}$ at $Re$ = 3000 and $Re$ = 200000.}
  \label{66677}
  \end{center}
\end{table}

Substituting (\ref{New_U_profile}) into (\ref{1}) yields the perturbations equations:

\begin{equation}
\frac{\partial u}{\partial t} + (U - C) \frac{\partial u}{\partial x'} + \frac{\mathrm{d}U}{\mathrm{d}y}\cdot v + \frac{\partial p}{\partial x'} - \frac{1}{Re} \nabla^2 u = \frac{\partial}{\partial x'}(u\widetilde{u^{\mathrm{2D}}}) + \frac{\partial}{\partial y}(v\widetilde{u^{\mathrm{2D}}} + u\widetilde{v^{\mathrm{2D}}}), 
\label{2345}
\end{equation}

\begin{equation}
\frac{\partial v}{\partial t} + (U - C) \frac{\partial v}{\partial x'} + \frac{\partial p}{\partial y} - \frac{1}{Re} \nabla^2 v =\frac{\partial}{\partial x'}(u\widetilde{v^{\mathrm{2D}}} + v\widetilde{u^{\mathrm{2D}}}) + \frac{\partial}{\partial y}(v\widetilde{v^{\mathrm{2D}}}).
\label{6789}
\end{equation}
According to Floquet theory \citep{22}, the solution of such equations with periodic coefficients can be expanded in Fourier series as:

\begin{equation}
u(x',y,t) = e^{\lambda t} \sum_m \tilde{u}_m(y)e^{i(m\alpha + \gamma_i)x'}
\label{577}
\end{equation}
\begin{equation}
v(x',y,t) = e^{\lambda t} \sum_m \tilde{v}_m(y)e^{i(m\alpha + \gamma_i)x'}
\label{588}
\end{equation}
where $m \in \mathbb{Z}$ and $\lambda$ denotes the complex time frequency. The real part of $\lambda$ is $\lambda_r$. When $\lambda_r>0$, it indicates that the flow is linearly unstable, otherwise it is linearly stable. $\gamma_i$ is the Floquet parameter, it has three selection methods corresponding to three types of secondary instability modes: (1) fundamental modes: $\gamma_i=0$; (2) subharmonic modes: $\gamma_i=\alpha/2$; (3) detuned modes: $0<\gamma_i<\alpha/2$.

Substituting (\ref{577}) and (\ref{588}) into (\ref{2345}) and (\ref{6789}) yields the modal equations:

\begin{align}
\lambda \tilde{u}_m + i \alpha_m (U - C) \tilde{u}_m + \tilde{v}_m DU + i \alpha_m \tilde{p} 
    - \frac{1}{Re} (D^2 - \alpha_m^2) \tilde{u}_m &= N_u, \label{eq:um} \\
\lambda \tilde{v}_m + i \alpha_m (U - C) \tilde{v}_m + D \tilde{p} 
    - \frac{1}{Re} (D^2 - \alpha_m^2) \tilde{v}_m &= N_v, \label{eq:vm} \\
i \alpha_m \tilde{u}_m + D \tilde{v}_m &= 0. \label{eq:cont}
\end{align}
where

\begin{subequations}
\label{eq:system1}
\begin{align}
D &= \frac{\mathrm{d}}{\mathrm{d}y}, \label{eq:D}\\
N_u &=  i \alpha_m \tilde{u}_{m\pm1} u^{\mathrm{2D}} + D (\tilde{v}_{m\pm1} u^{\mathrm{2D}} + \tilde{u}_{m\pm1} v^{\mathrm{2D}}), \label{eq:Nu}\\
N_v &=  i \alpha_m (\tilde{u}_{m\pm1} v^{\mathrm{2D}} + \tilde{v}_{m\pm1} u^{\mathrm{2D}}) + D (\tilde{v}_{m\pm1} v^{\mathrm{2D}}), \label{eq:Nv}\\
\alpha_m &= m\alpha + \gamma_i, \label{eq:alpha}\\
\tilde{u}_{m\pm 1} &= \tilde{u}_{m+1} + \tilde{u}_{m-1}. \label{eq:utilde}
\end{align}
\end{subequations}

\subsection{Truncation Method}
Firstly, by substituting the continuity equation (\ref{eq:cont}) and the streamwise momentum equation (\ref{eq:um}) into the normal momentum equation (\ref{eq:vm}), the pressure $\tilde{p}$ and streamwise perturbation velocity $\tilde{u}_m$ can be eliminated, yielding an equation solely for the normal perturbation velocity $\tilde{v}_m$:

\begin{equation}
\mathcal{L}\tilde{v}_{m} = \mathcal{L}_1 \tilde{v}_{m-1} + \mathcal{L}_2 \tilde{v}_{m+1}
\end{equation}
where 

\begin{subequations}
\label{eq:system2}
\begin{align}
\mathcal{L} &= \left[ \lambda + i\alpha_m(U - C) - \frac{1}{Re}(D^2 - \alpha_m^2) \right](D^2 - \alpha_m^2) - i\alpha_m D^2U  \\
\mathcal{L}_1 &= 
\begin{aligned}[t]
& \bigl( D^{2} u^{2D} + \alpha_{m}^{2} u^{2D} - i \alpha_{m} D v^{2D} \bigr) \\
& + \left[ \left(2 - \frac{\alpha_m}{\alpha_m - \alpha}\right) D u^{2D} - \frac{1}{i(\alpha_m - \alpha)} D^{2} v^{2D} + i\left( \frac{\alpha_m^{2}}{\alpha_m - \alpha} - \alpha_m \right) v^{2D} \right] D \\
& + \left[ \left(1 - \frac{\alpha_m}{\alpha_m - \alpha}\right) u^{2D} - \frac{2}{i(\alpha_m - \alpha)} D v^{2D} \right] D^{2} \\
& + \left[ -\frac{1}{i(\alpha_m - \alpha)} v^{2D} \right] D^{3}
\end{aligned}  \\
\mathcal{L}_2 &= 
\begin{aligned}[t]
& \bigl( D^{2} u^{2D} + \alpha_{m}^{2} u^{2D} - i \alpha_{m} D v^{2D} \bigr) \\
& + \left[ \left(2 - \frac{\alpha_m}{\alpha_m + \alpha}\right) D u^{2D} - \frac{1}{i(\alpha_m + \alpha)} D^{2} v^{2D} + i\left( \frac{\alpha_m^{2}}{\alpha_m + \alpha} - \alpha_m \right) v^{2D} \right] D \\
& + \left[ \left(1 - \frac{\alpha_m}{\alpha_m + \alpha}\right) u^{2D} - \frac{2}{i(\alpha_m + \alpha)} D v^{2D} \right] D^{2} \\
& + \left[ -\frac{1}{i(\alpha_m + \alpha)} v^{2D} \right] D^{3}
\end{aligned} 
\end{align}
\end{subequations}

Secondly, for the secondary instability modes, the subharmonic modes were chosen to determine the phase difference \(\gamma_i = \alpha / 2\) in (\ref{588}) \citep{23}. Subsequently, the normal perturbation velocity takes the following form:

\begin{equation}
v(x',y,t) = e^{\lambda t} \sum_m \tilde{v}_m(y)e^{i(m\alpha + \alpha/2)x'}.
\end{equation}
Note that, since $m \in \mathbb{Z}$, the secondary instability equations would be infinite-dimensional. Therefore, the system must be truncated to facilitate subsequent matrix calculations. In the case of subharmonic secondary instability, the truncation scheme can be chosen following \cite{24} and \cite{1} by truncating at \( m = 0 \) and \( m = -1 \), ensuring that the wavenumbers corresponding to the two modes \( \tilde{v}_0 \) and \( \tilde{v}_{-1} \) are exactly opposite: 

\begin{equation}
\alpha_0 = \frac{\alpha}{2}, \quad \alpha_{-1} = -\frac{\alpha}{2}
\end{equation}

At this stage, we can transform the perturbation equations into a matrix eigenvalue problem similar to linear stability analysis:

\begin{equation}
\lambda \left( \begin{array}{cc} L_1 & 0 \\ 0 & L_4 \end{array} \right) \left( \begin{array}{c} \tilde{v}_0 \\ \tilde{v}_{-1} \end{array} \right) = \left( \begin{array}{cc} L_2 & L_3 \\ L_5 & L_6 \end{array} \right) \left( \begin{array}{c} \tilde{v}_0 \\ \tilde{v}_{-1} \end{array} \right)
\end{equation}
where the differential operators are defined as:

\begin{subequations}
\label{eq:system3}
\begin{align}
L_1 &= \frac{2i}{\alpha} D^2 - \frac{\alpha i}{2} \\
L_2 &= 
\begin{aligned}[t]
& (U - C)D^2 - \frac{\alpha}{2}(U - C) - D^2U + \frac{2i}{\alpha Re} D^4 - \frac{\alpha i}{2Re} D^2 \\
& - 2u^{2D} D^2 - 3Du^{2D} D - D^2 u^{2D} + \frac{2i}{\alpha} v^{2D} D^3
\end{aligned} \\
L_3 &= 
\begin{aligned}[t]
& \frac{4i}{\alpha} Dv^{2D} D^2 + \frac{2i}{\alpha} D^2 v^{2D} D + \frac{\alpha i}{2} v^{2D} D - \frac{\alpha^2}{4} u^{2D} \\
& + \frac{\alpha i}{2} Dv^{2D} + \frac{\alpha i}{2} v^{2D} D
\end{aligned} \\
L_4 &= -\frac{2i}{\alpha} D^2 + \frac{\alpha i}{2} \\
L_5 &= 
\begin{aligned}[t]
& -\frac{4i}{\alpha} Dv^{2D} D^2 - \frac{2i}{\alpha} D^2 v^{2D} D - \frac{\alpha i}{2} v^{2D} D - \frac{\alpha^2}{4} u^{2D} \\
& - \frac{\alpha i}{2} Dv^{2D} - \frac{\alpha i}{2} v^{2D} D
\end{aligned} \\
L_6 &= 
\begin{aligned}[t]
& (U - C)D^2 + \frac{\alpha}{2}(U - C) - D^2U - \frac{2i}{\alpha Re} D^4 + \frac{5\alpha i}{2Re} D^2 - \frac{\alpha^3 i}{2Re} \\
& - 2u^{2D} D^2 - 3Du^{2D} D - D^2 u^{2D} - \frac{2i}{\alpha} v^{2D} D^3
\end{aligned}
\end{align}
\end{subequations}

\subsection{Eigenvalue Spectrum }
A generalized eigenvalue problem is obtained:

\begin{align}
\mathbf{A} \tilde{\mathbf{q}} = \lambda \mathbf{B} \tilde{\mathbf{q}},
\end{align}
where

\begin{equation}
\tilde{\mathbf{q}}= \begin{pmatrix} \tilde{v}_0 \\ \tilde{v}_{-1} \end{pmatrix}, \quad 
\mathbf{A} = \begin{pmatrix} L_2 & L_3 \\ L_5 & L_6 \end{pmatrix}, \quad 
\mathbf{B} = \begin{pmatrix} L_1 & 0 \\ 0 & L_4 \end{pmatrix}
\end{equation}
and

\begin{equation}
\tilde{v}_0 = D\tilde{v}_0 = \tilde{v}_{-1} = D\tilde{v}_{-1} = 0   \text{ at } y = \pm 1
\end{equation}

In order to obtain the eigenvalue spectrum numerically, it is necessary to discretize the differential matrix. The specific discretization method can be found in Appendix~\ref{Discrete Method of Differential Matrix}.

\begin{figure}[h]
    \centering
    \includegraphics[width=0.6\textwidth]{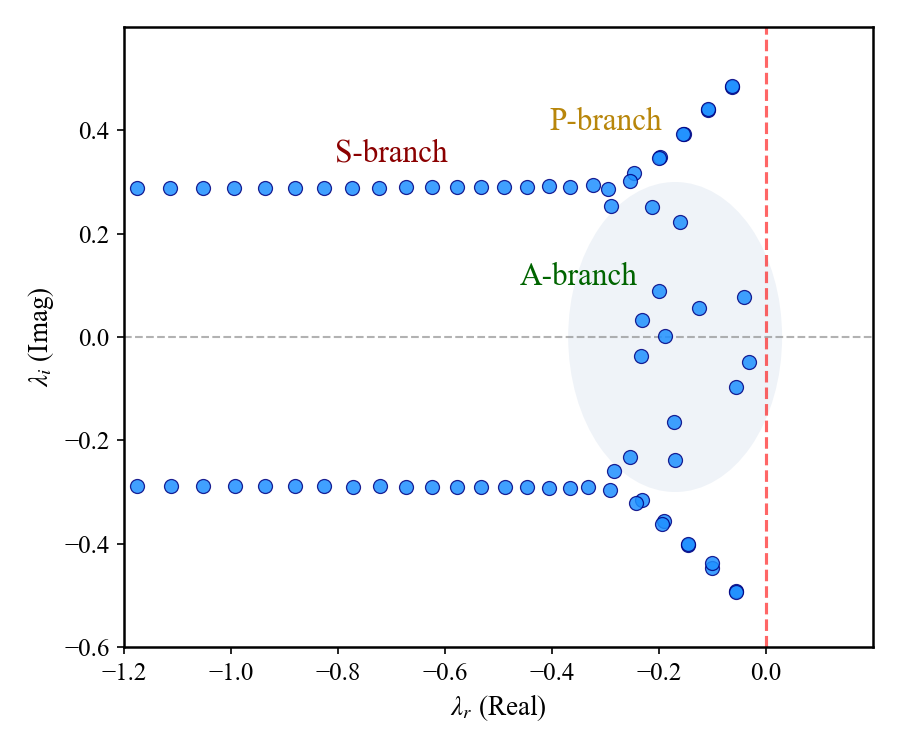}
    \caption{Eigenvalue spectrum at $Re = 3000$ by scaling the traveling wave term $u^{2D}$ and $v^{2D}$ with a small parameter $\epsilon\sim10^{-6}$}
    \label{fig:27}
\end{figure}

Before formally calculating the eigenvalue spectrum of secondary instability, to ensure the accuracy of the code, the traveling wave term $u^{2D}$ and $v^{2D}$ in the matrix equation are multiplied by a small parameter $\epsilon\sim10^{-6}$ to eliminate their influence. The resulting eigenvalue distribution, shown in figure~\ref{fig:27}, exhibits the classic Y-shaped pattern, confirming the correctness of the code.

\begin{figure}[h]
    \centering
    \hspace*{0cm}
    \includegraphics[width=1\textwidth]{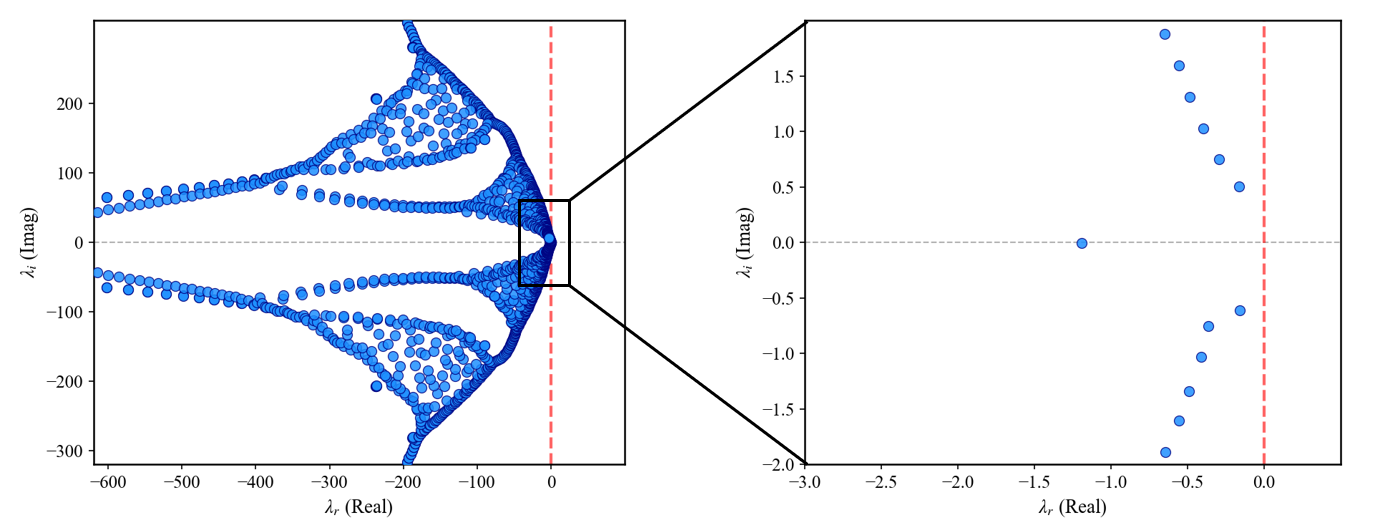}
    \caption{Eigenvalue spectrum at $Re$ = 3000}
    \label{fig:4}
\end{figure}

\begin{figure}[h]
    \centering
    \hspace*{-0.2cm}
    \includegraphics[width=0.98\textwidth]{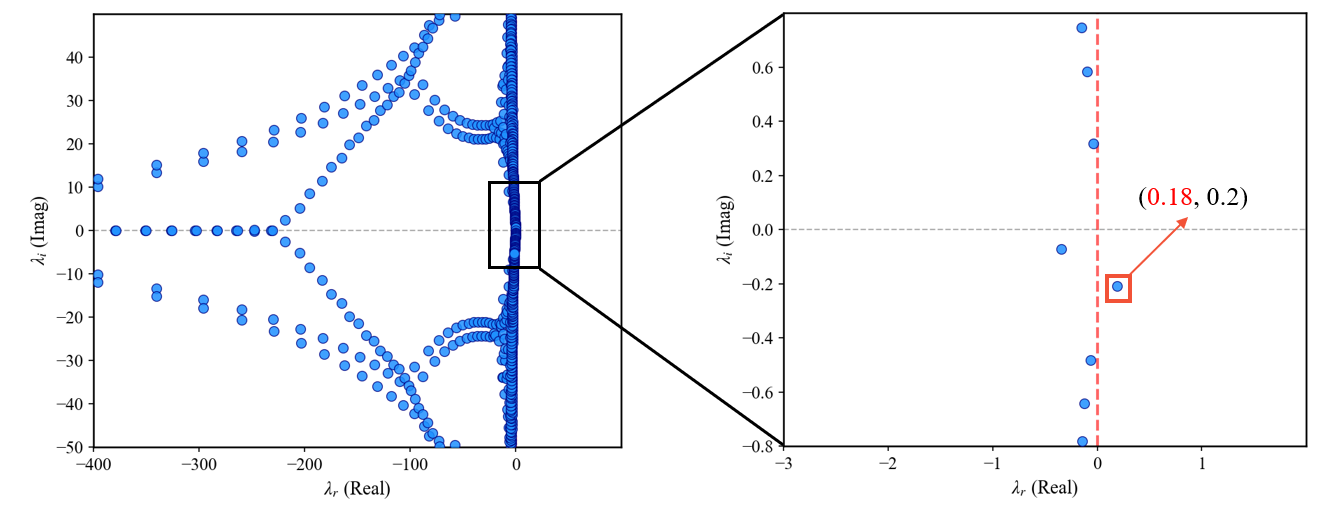}
    \caption{Eigenvalue spectrum at $Re$ = 200000}
    \label{fig:5}
\end{figure}

The computed eigenvalue spectrum is shown in figure~\ref{fig:4} and figure~\ref{fig:5}. The horizontal axis represents the real part of teh eigenvalues, and the vertical axis represents the imaginary part.

It can be seen that at $Re = 3000$, the eigenvalue spectrum of the linear instability for the large-scale traveling wavy structure does not possess any eigenvalues with a positive real part (figure~\ref{fig:4}). This proves that the structure is linearly stable, which is consistent with the DNS results. Conversely, at $Re = 200000$, the eigenvalue spectrum does possess one eigenvalue with a positive real part (figure~\ref{fig:5}). This indicates that the structure is linearly unstable, referred to as the subharmonic torsional mode. Therefore, the turbulence characteristics observed in the DNS results can be explained from the perspective of linear instability.

\begin{figure}[h]
    \centering
    \begin{subfigure}{0.45\textwidth}
        \centering
        \begin{overpic}[width=\linewidth, trim=-10pt 0 -10pt 0, clip]{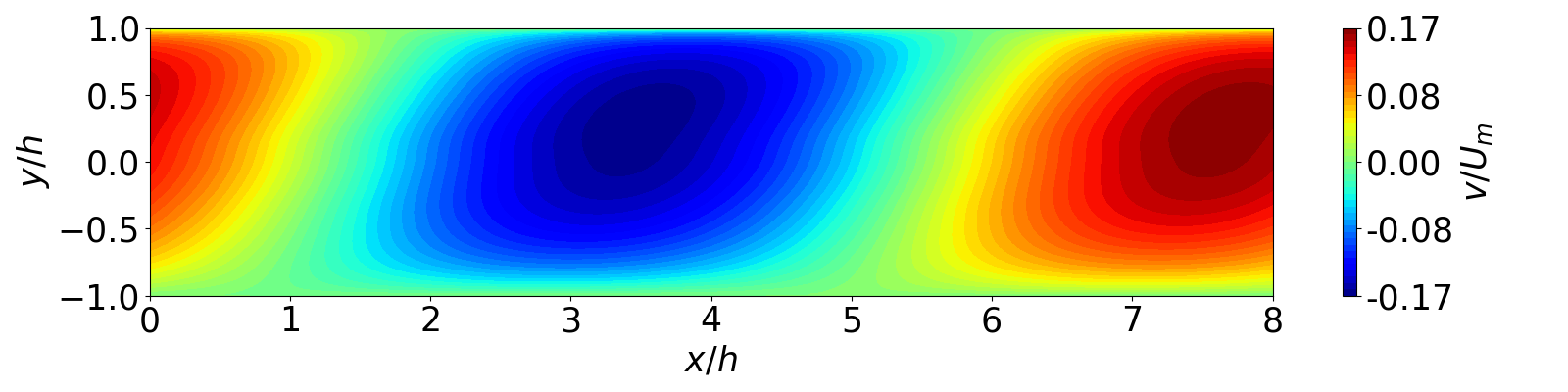}
            \put(-5,20){\small(a)}
        \end{overpic}
    \end{subfigure}
    \begin{subfigure}{0.45\textwidth}
        \centering
        \begin{overpic}[width=\linewidth, trim=-10pt 0 -10pt 0, clip]{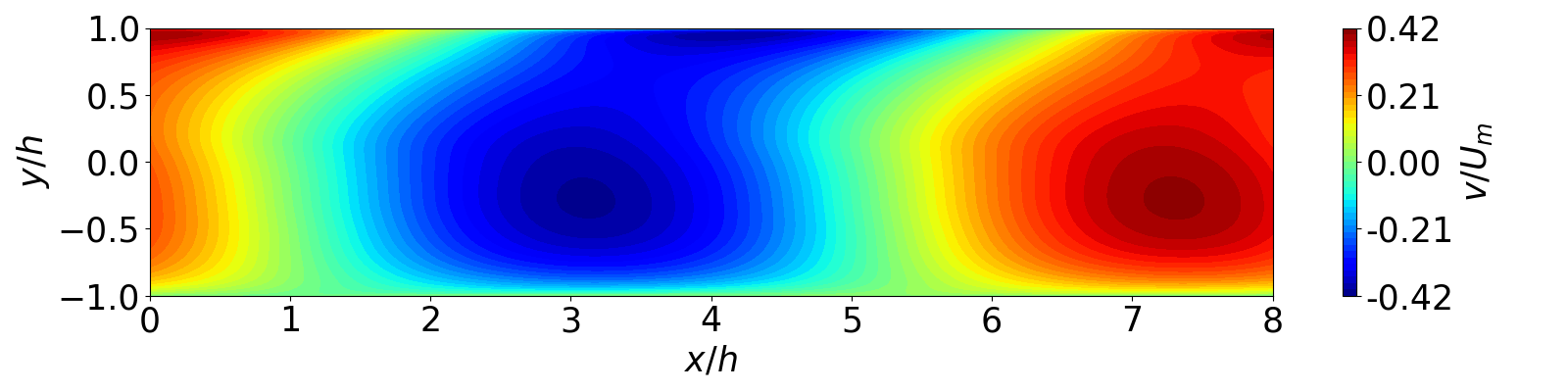}
            \put(-5,20){\small(b)}
        \end{overpic}
    \end{subfigure}\\
    \begin{subfigure}{0.45\textwidth}
        \centering
        \begin{overpic}[width=\linewidth, trim=-10pt 0 -10pt 0, clip]{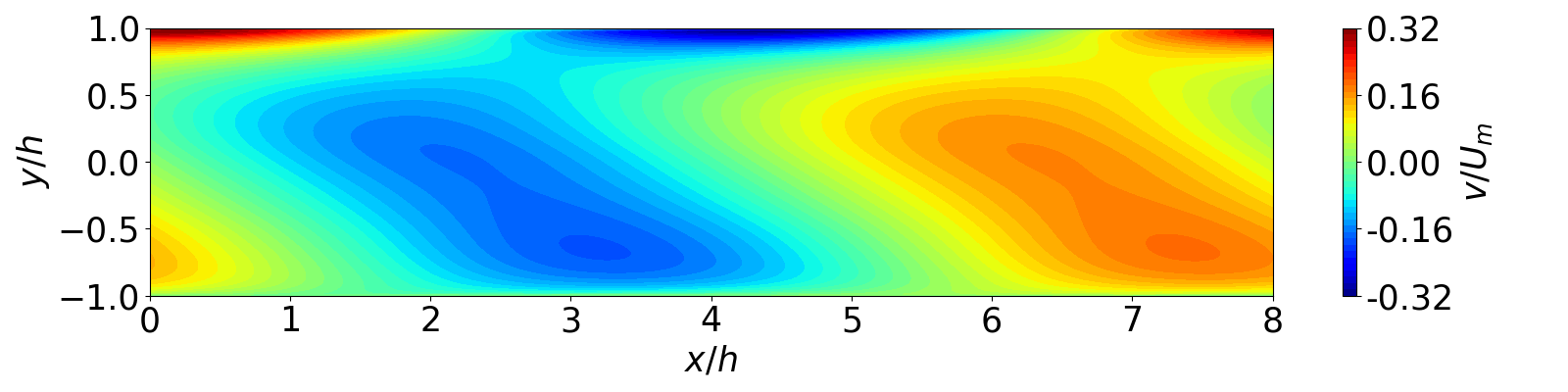}
            \put(-5,20){\small(c)}
        \end{overpic}
    \end{subfigure}
    \begin{subfigure}{0.45\textwidth}
        \centering
        \begin{overpic}[width=\linewidth, trim=-10pt 0 -10pt 0, clip]{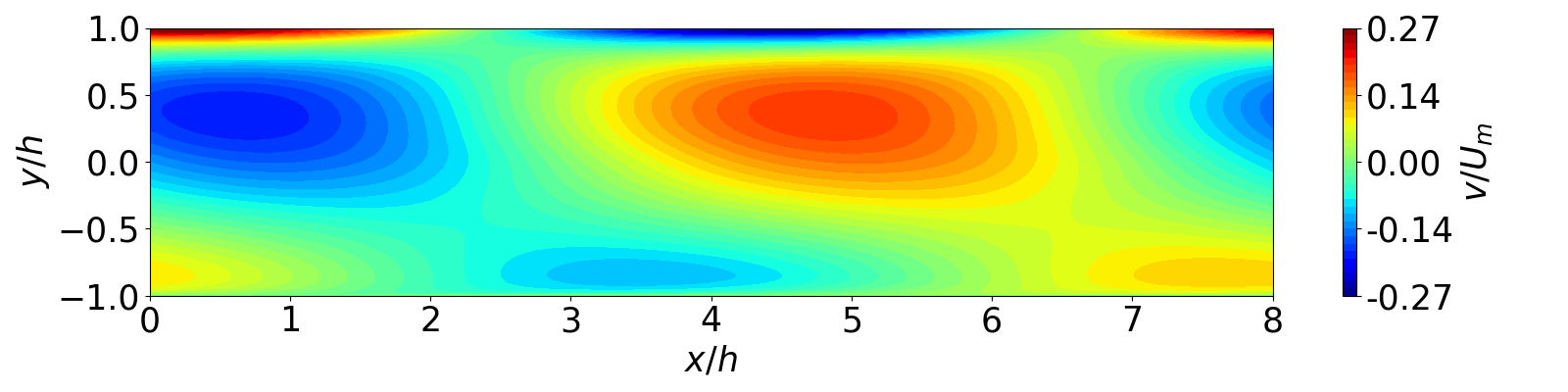}
            \put(-5,20){\small(d)}
        \end{overpic}
    \end{subfigure}
    \begin{subfigure}{0.45\textwidth}
        \centering
        \begin{overpic}[width=\linewidth, trim=-10pt 0 -10pt 0, clip]{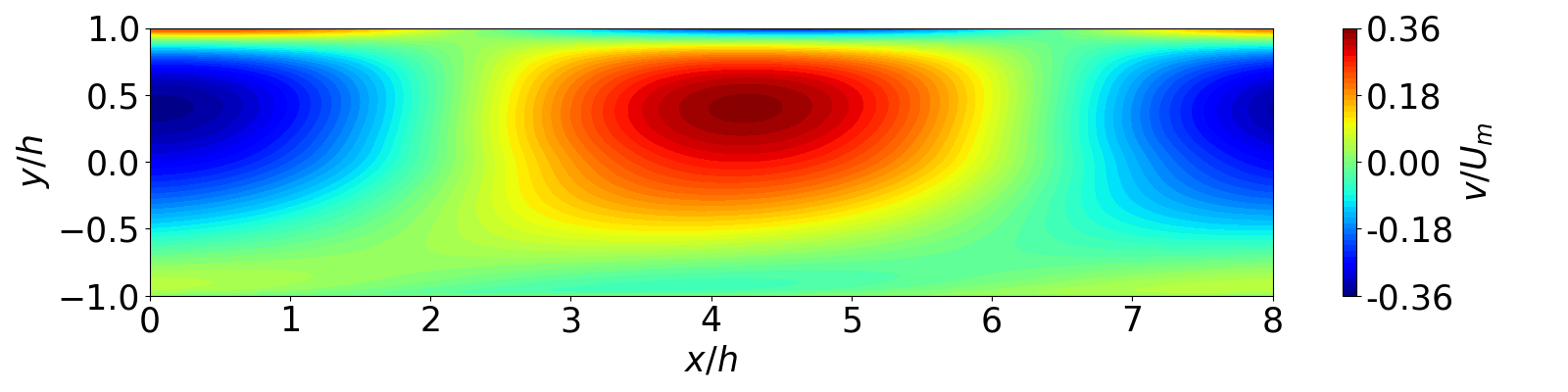}
            \put(-5,20){\small(e)}
        \end{overpic}
    \end{subfigure}
    \begin{subfigure}{0.45\textwidth}
        \centering
        \begin{overpic}[width=\linewidth, trim=-10pt 0 -10pt 0, clip]{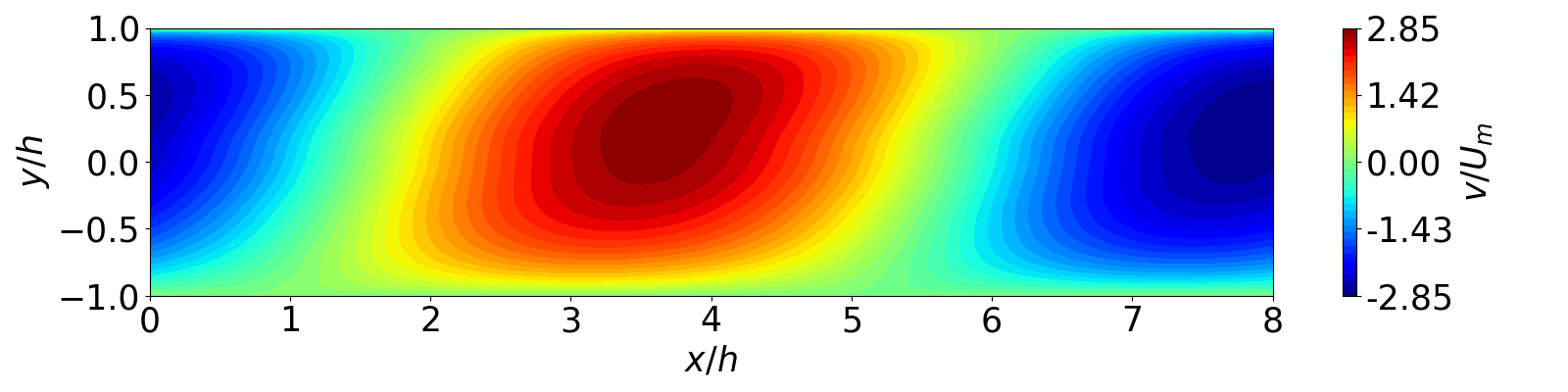}
            \put(-5,20){\small(f)}
        \end{overpic}
    \end{subfigure}
    \caption{Evolution of the dimensionless normal velocity \(v/U_m\) of the unstable mode at $Re=200000$ over a period of nondimensional time $\hat{t}=t \cdot U_m/h$. (a) $\hat{t}=0$; (b) $\hat{t}=9$; (c) $\hat{t}=9.7$; (d) $\hat{t}=10.8$; (e) $\hat{t}=11.3$; (f) $\hat{t}=15$.}
    \label{fig:unstable_mode_evolution}
\end{figure}

\begin{figure}[h]
    \centering
    \begin{subfigure}{0.45\textwidth}
        \centering
        \begin{overpic}[width=\linewidth, trim=-10pt 0 -10pt 0, clip]{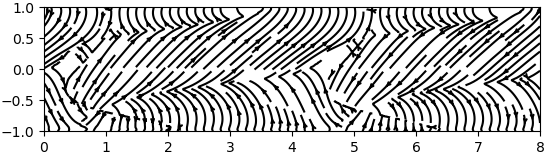}
            \put(-5,20){\small(a)}
        \end{overpic}
    \end{subfigure}
    \begin{subfigure}{0.45\textwidth}
        \centering
        \begin{overpic}[width=\linewidth, trim=-10pt 0 -10pt 0, clip]{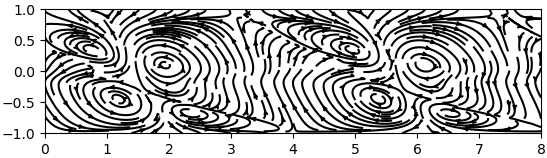}
            \put(-4,20){\small(b)}
        \end{overpic}
    \end{subfigure}\\
    \begin{subfigure}{0.45\textwidth}
        \centering
        \begin{overpic}[width=\linewidth, trim=-10pt 0 -10pt 0, clip]{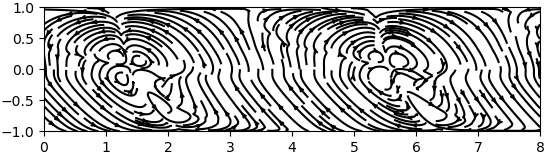}
            \put(-5,20){\small(c)}
        \end{overpic}
    \end{subfigure}
    \begin{subfigure}{0.45\textwidth}
        \centering
        \begin{overpic}[width=\linewidth, trim=-10pt 0 -10pt 0, clip]{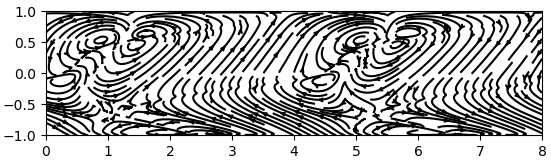}
            \put(-4,20){\small(d)}
        \end{overpic}
    \end{subfigure}
    \begin{subfigure}{0.45\textwidth}
        \centering
        \begin{overpic}[width=\linewidth, trim=-10pt 0 -10pt 0, clip]{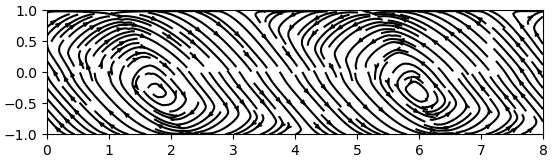}
            \put(-5,20){\small(e)}
        \end{overpic}
    \end{subfigure}
    \begin{subfigure}{0.45\textwidth}
        \centering
        \begin{overpic}[width=\linewidth, trim=-10pt 0 -10pt 0, clip]{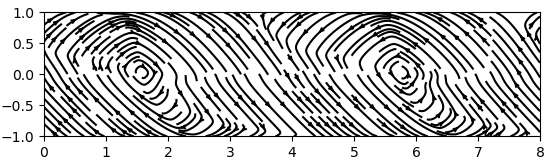}
            \put(-4,20){\small(f)}
        \end{overpic}
    \end{subfigure}
    \caption{Temporal evolution of the streamlines of the unstable mode at $Re=200000$ over a period of nondimensional time $\hat{t}=t \cdot U_m/h$. (a) $\hat{t}=0$; (b) $\hat{t}=9$; (c) $\hat{t}=9.7$; (d) $\hat{t}=10.8$; (e) $\hat{t}=11.3$; (f) $\hat{t}=15$.}
    \label{fig:streamlines}
\end{figure}

In summary, at $Re=3000$, the large-scale traveling wavy structure formed by adding perturbations and undergoing nonlinear saturation in the 2D channel is linearly stable, and over time, this structure remains stable in the flow field. At $Re=200000$, this large-scale traveling wavy structure is linearly unstable, with a growth rate of $\lambda_r=0.18$. The only unstable mode found is presented in Figure~\ref{fig:unstable_mode_evolution}. Similarly, $\tilde{u}_m$ can be obtained from the formula $i \alpha_m \tilde{u}_m + D \tilde{v}_m = 0$, and the streamlines of this unstable mode can then be plotted in Figure~\ref{fig:streamlines}.

In Figure~\ref{fig:unstable_mode_evolution}, the temporal evolution of the unstable mode is presented over the nondimensional time $\hat{t}=t\cdot U_m/h$, covering the range from $\hat{t}=0$ to $\hat{t}=15$. A clear progression from a regular traveling wavy structure to a complex multi-scale morphology can be observed. At the initial instant ($\hat{t}=0$), the perturbation exhibits a regular large-scale wavy pattern. Then, at $\hat{t}=9$, the wavy structure begins to show visible deformation and loses its symmetry. In the intermediate stage from $\hat{t}=9.7$ to $\hat{t}=10.8$, the upper and lower parts of the wave gradually split from the main wave into three rows of traveling wavy structures with different scales. Ultimately, from $\hat{t}=11.3$ to $\hat{t}=15$, the traveling wavy structures that split from the upper and lower ends gradually disappear, while the main wave expands until it occupies the original spatial scale and exhibits a phase opposite to that of the initial wave. Throughout the whole process, the perturbation amplitude continues to grow exponentially under the influence of the positive real part of the eigenvalue $\lambda_r=0.18>0$.

From a mathematical perspective, this evolution is governed by the modal decomposition and coupling mechanisms inherent to secondary instability theory. The perturbation velocity field can be expressed via the Floquet expansion

\begin{equation}
v(x',y,t) = \mathrm{Re}\left\{ e^{\lambda t} \left[ \tilde{v}_0(y) e^{i\frac{\alpha}{2}x'} + \tilde{v}_{-1}(y) e^{-i\frac{\alpha}{2}x'} \right] \right\},
\end{equation}
where $\lambda = \lambda_r + i\lambda_i$, $\tilde{v}_0(y) = a_0(y) + i b_0(y)$, and $\tilde{v}_{-1}(y) = a_{-1}(y) + i b_{-1}(y)$. Therefore, $v(x,y,t)$ can be written in expanded form as

\begin{equation}
\begin{aligned}
v(x, y, t) = e^{\lambda_r t} \Big[ 
& a_0(y) \cos\left(\frac{\alpha}{2}x - \frac{\alpha}{2}Ct + \lambda_i t\right) 
- b_0(y) \sin\left(\frac{\alpha}{2}x - \frac{\alpha}{2}Ct + \lambda_i t\right) \\
& + a_{-1}(y) \cos\left(\frac{\alpha}{2}x - \frac{\alpha}{2}Ct - \lambda_i t\right) 
+ b_{-1}(y) \sin\left(\frac{\alpha}{2}x - \frac{\alpha}{2}Ct - \lambda_i t\right) \Big].
\end{aligned}
\end{equation}

From the real-valued expression derived from the Floquet expansion, it can be seen that the waveform is composed of four sinusoidal waves sharing the same spatial frequency but differing in their temporal oscillation phases. Their wall-normal coefficients \(a_0(y), b_0(y), a_{-1}(y), b_{-1}(y)\) vary with \(y\) and are distinct from one another, leading to different weights of each mode at different heights. This difference in weights, combined with the temporal phase differences, causes the wave crests and troughs to shift along the streamwise direction in a wall-normal-dependent manner, resulting in waveform distortion. When two of the sinusoidal waves dominate in certain wall-normal regions, the upper and lower parts of the flow begin to exhibit spatially misaligned wave trains, a phenomenon referred to as ``splitting''. Ultimately, due to the inhomogeneous wall-normal distribution, the superimposed waveform visually displays multi-scale structures, although it fundamentally consists of only two Fourier modes with a single spatial frequency \(\alpha/2\).

\section{Conclusion}
\label{conclusion}
In this study, we have  investigated the linear stability of large-scale traveling wavy structures arising from the nonlinear saturation of perturbations in two-dimensional channel flows. By combining direct numerical simulations (DNS), singular value decomposition (SVD), and Floquet-based secondary instability analysis, we examined the stability characteristics of these coherent structures at two Reynolds numbers, \(Re = 3000\) and \(Re = 200000\), which lie on opposite sides of the transitional Reynolds number \(Re \approx 10000\) reported by \cite{chen2024}.

At \(Re = 3000\), the large-scale traveling wavy structure is linearly stable. The eigenvalue spectrum contains no eigenvalues with positive real parts, indicating that any infinitesimal perturbations imposed on the coherent structure will decay. This linear stability result is fully consistent with the DNS observations, which show a flow field devoid of small-scale turbulent fluctuations and characterized by a persistent, regular wavy pattern. The energy distribution across the SVD modes further supports this conclusion: even low-rank approximations retain a high proportion of the total kinetic energy, confirming the laminar nature of the flow.

At \(Re = 200000\), a markedly different behaviour is observed. The raw DNS data exhibit clear turbulent features, including small-scale velocity fluctuations that contaminate the large-scale coherent structure. To isolate the underlying traveling wavy structure, we applied SVD-based filtering, retaining a rank-2 approximation for the streamwise velocity and a rank-1 approximation for the wall-normal velocity. This procedure effectively removes low-energy small-scale fluctuations while preserving the dominant coherent pattern. Subsequent secondary instability analysis of the filtered flow field uncovers a distinct unstable eigenmode with a positive real part \(\lambda_r = 0.18 > 0\), providing direct evidence that the large-scale traveling wavy structure is linearly unstable at this Reynolds number. Temporal reconstruction of this unstable mode from the Floquet expansion reveals a rich dynamical process: starting from a regular wavy pattern, the perturbation undergoes gradual deformation, loss of symmetry, and eventual spatial splitting into multiple wave trains of differing scales, followed by a phase-reversed reconstruction of the main wave. Notably, despite the apparent emergence of multi-scale structures in physical space, the mathematical decomposition demonstrates that the instability is fundamentally governed by only two Fourier modes sharing a single spatial frequency \(\alpha/2\), with the complex waveform arising from the inhomogeneous wall-normal distribution of the modal coefficients and their distinct temporal phases.

These findings establish a direct causal link between the linear instability of the large-scale traveling wavy structure and the onset of turbulent characteristics in two-dimensional channel flows. In particular, they demonstrate that the transition to turbulence at sufficiently high Reynolds numbers can occur without invoking three-dimensional perturbations --- a scenario that contrasts sharply with classical secondary instability theory in three-dimensional shear flows \citep{17,18}. Instead, the two-dimensional coherent structure itself becomes intrinsically unstable, providing a self-contained pathway from laminar-like organised motion to a chaotic, multi-scale turbulent state.

From a broader perspective, this work contributes to the growing understanding of two-dimensional wall-bounded turbulence by shifting the focus from the statistical properties of the turbulent cascade to the dynamical stability of the coherent structures that emerge from the inverse energy cascade. The methodology developed here --- combining DNS, SVD-based filtering, and Floquet stability analysis --- offers a general framework for interrogating the stability of large-scale coherent structures in other two-dimensional flow configurations, such as soap films, geophysical flows, and microfluidic devices. Future work may extend this analysis to a wider range of Reynolds numbers to map out the neutral stability curve and identify the critical conditions under which the large-scale traveling wavy structure becomes unstable. Furthermore, incorporating weakly nonlinear theory could elucidate the saturation mechanisms that ultimately bound the growth of the unstable mode and determine the final turbulent state. Finally, the connection between the linear instability identified here and the statistical scaling laws observed in two-dimensional wall turbulence, such as the \(C_f \sim Re^{-1/2}\) friction law, remains an intriguing avenue for future investigation.

\begin{bmhead}[Acknowledgments.]
X. Chen appreciates the funding support provided by the National Key Research and Development Program of China (2022YFF0610805), the National Natural Science Foundation of China (Grant Nos. 12072012 and 92252201).
\end{bmhead}

\begin{bmhead}[Author declarations conflict of interest.]
The authors have no conficts to disclose.
\end{bmhead}

\begin{bmhead}[Data availability.]
The data that support the findings of this study are available from the corresponding author upon reasonable request.
\end{bmhead}

\begin{appen}

\section{Numerical Setup}\label{Numerical Setup}
The incompressible 2D Navier-Stokes equations (with the density $\rho$ absorbed into the pressure $\mathfrak{p}$),
\begin{equation}
\partial_t \mathfrak{u}_i + \partial_j \mathfrak{u}_i \mathfrak{u}_j = -\partial_i \mathfrak{p} + \nu \partial_j \partial_j \mathfrak{u}_i
\end{equation}
\begin{equation}
\partial_i \mathfrak{u_i} = 0
\end{equation}
are solved using our finite difference code which has been well validated for 3D \citep{xie2021low} and 2D \citep{chen2024} channel flows. 
Simulation are conducted in the domain shown in figure \ref{fig:60}, with periodic boundary conditions in the streamwise direction and no-slip boundary conditions at both the top and bottom walls. 
A second-order centered finite difference method with staggered grids is used for spatial discretization, and the second-order Runge-Kutta scheme is employed for time advancement. 
Fast Fourier transforms are employed to solve the pressure Poisson equation. 
The method used to sustain turbulence is based on a constant flow rate, while details on the parallelization are given by \cite{xie2021low}. 
A summary of the simulations is shown in Table~\ref{HHHHH}.

\begin{table}
  \begin{center}
\def~{\hphantom{0}}
  \begin{tabular}{lcccc}
    $Re$ & $N_x \times N_y$ & $\Delta x^+$ & $\Delta y^+$ & $\Delta t/(h/U_m)$   \\[3pt]
       3000 & $1024 \times 320$ & 0.99 & 0.05--1.58 & $2 \times 10^{-4}$  \\
     200000 & $4096 \times 2048$ & 5.55 & 0.18--5.52 & $5 \times 10^{-5}$ \\
  \end{tabular}
  \caption{Parameters of the simulations. $N_x$ is the number of grid cells in the streamwise direction and $N_y$ is that in the wall-normal direction. $\Delta x^+$ and $\Delta y^+$ represent the grid spacings normalized by viscous units. $\Delta t/(h/U_m)$ represents the time step $\Delta t$ normalized by $h/U_m$, where \( h \) is the half height of the channel, and \( U_m \) is the bulk velocity.}
  \label{HHHHH}
  \end{center}
\end{table}

\section{Derivation of SVD}
\label{Complete Derivation of SVD}
For any real \( m \times n \) matrix \( A \), there exist orthogonal matrices \( U \), \( V \), and a diagonal matrix \( \Sigma \), such that:
\begin{equation}
A = U \Sigma V^T
\end{equation}
\( U \) is an \( m \times m \) orthogonal matrix (\( U^T U = I_m \)), whose column vectors are called the left singular vectors. \( V \) is an \( n \times n \) orthogonal matrix (\( V^T V = I_n \)), whose column vectors are called the right singular vectors. \( \Sigma \) is an \( m \times n \) diagonal matrix of the form:
\begin{equation}
\Sigma =
\begin{pmatrix}
\sigma_1 & 0 & \cdots & 0 & 0 & \cdots & 0 \\
0 & \sigma_2 & \cdots & 0 & 0 & \cdots & 0 \\
\vdots & \vdots & \ddots & \vdots & \vdots & \vdots & \vdots \\
0 & 0 & \cdots & \sigma_r & 0 & \cdots & 0 \\
0 & 0 & \cdots & 0 & 0 & \cdots & 0 \\
\vdots & \vdots & \vdots & \vdots & \vdots & \vdots & \vdots \\
0 & 0 & \cdots & 0 & 0 & \cdots & 0
\end{pmatrix}
\end{equation}
where \( \sigma_1 \geq \sigma_2 \geq \cdots \geq \sigma_r > 0 \), and \( r = \text{rank}(A) \). These \( \sigma_i \) are called the singular values.

\( M = A^T A \), is an \( n \times n \) symmetric matrix. And \( M \) is positive semidefinite:\\  
\begin{equation}
x^T M x = (Ax)^T (Ax) = \|Ax\|^2 \geq 0 \quad \text{for any } x \in \mathbb{R}^n
\end{equation}
Therefore, the eigenvalues of \( M \) are all real and nonnegative, denoted as:
\begin{equation}
\lambda_1 \geq \lambda_2 \geq \cdots \geq \lambda_r > 0, \quad \lambda_{r+1} = \cdots = \lambda_n = 0
\end{equation}
where \( r = \text{rank}(A) \), since  
\begin{equation}
\text{rank}(A^T A) = \text{rank}(A)
\end{equation}

Perform orthogonal diagonalization of \( M \):
\begin{equation}
M = V \Lambda V^T
\end{equation}
where \( V \) is an \( n \times n \) orthogonal matrix whose column vectors \( v_1, v_2, \ldots, v_n \) are the orthonormal eigenvectors of \( M \), corresponding to eigenvalues \( \lambda_i \).
\begin{equation}
\Lambda = \text{diag}(\lambda_1, \ldots, \lambda_n)
\end{equation}

Let:
\begin{equation}
\sigma_i = \sqrt{\lambda_i}, \quad i = 1, \ldots, n
\end{equation}
then \(\sigma_1 \geq \cdots \geq \sigma_r > 0\), \(\sigma_{r+1} = \cdots = \sigma_n = 0\).

$\Sigma$ is an \( m \times n \) matrix whose first \( r \) diagonal entries are \( \sigma_1, \ldots, \sigma_r \), and all other entries are zero.
More specifically:
\begin{equation}
\Sigma = 
\begin{bmatrix}
D & 0_{r \times (n-r)} \\
0_{(m-r) \times r} & 0_{(m-r) \times (n-r)}
\end{bmatrix}
\end{equation}
where \( D = \operatorname{diag}(\sigma_1, \ldots, \sigma_r) \).

Partition \( V \) column-wise as \( V = [v_1, v_2, \ldots, v_n] \).
For \( 1 \leq i \leq r \), define
\begin{equation}
u_i = \frac{1}{\sigma_i} A v_i
\end{equation} 
which can be derived directly from $AV = U\Sigma$. If \( r < m \), we need to supplement \( m - r \) orthonormal vectors \( u_{r+1}, \ldots, u_m \) such that together with \( u_1, \ldots, u_r \), they form an orthonormal basis of \( \mathbb{R}^m \) (which can be obtained via Gram–Schmidt or by arbitrary extension followed by orthogonalization). Let
\begin{equation}
U = [u_1, u_2, \ldots, u_r \mid u_{r+1}, \ldots, u_m]
\end{equation}
then \( U \) is an \( m \times m \) orthogonal matrix.

\section{Discretization Method of Differential Matrix}
\label{Discrete Method of Differential Matrix}
Select N+1 Chebyshev nodes in the normal direction $y \in [-1, 1]$ for discretization, which are determined by N+1 order Chebyshev polynomials and have the following form:
\begin{equation}
x_j = \cos\left(\frac{\pi j}{N}\right), \quad j = 0, 1, \ldots, N
\end{equation}

All velocity function terms $\mathcal{U}$ in the operator are only related to \(y\) and can be discretized into diagonal matrices:
\begin{equation}
\mathcal{U}_N = 
\begin{pmatrix}
\mathcal{U}(x_0) & 0 & \cdots & 0 \\
0 & \mathcal{U}(x_1) & \cdots & 0 \\
\vdots & \vdots & \ddots & \vdots \\
0 & 0 & \cdots & \mathcal{U}(x_{N})
\end{pmatrix}
\end{equation}
Differential operator $D= \frac{\mathrm{d}}{\mathrm{d}y}$ can be discretized into Chebyshev derivative matrix $D_N$:
\begin{subequations}
\label{eq:system4}
\begin{align}
(D_N)_{00} &= -\frac{2N^2 + 1}{6}\\
(D_N)_{NN} &= \frac{2N^2 + 1}{6} \label{eq:Nu}\\
(D_N)_{jj} &= \frac{-x_j}{2(1 - x_j^2)}, \quad j = 1, 2, 3, \ldots, N - 1 \\
(D_N)_{ij} &= \frac{c_i}{c_j} \frac{(-1)^{i+j}}{x_i - x_j}, \quad i \neq j, \quad i, j = 0, 1, 2, \ldots, N
\end{align}
\end{subequations}
where:
\begin{equation}
c_i = 
\begin{cases} 
2, & i = 0, N \\ 
1, & i = 1, 2, \ldots, N - 1 
\end{cases}
\end{equation}
For high-order differential operators $D^n$, it is only necessary to power the Chebyshev derivative matrices $D_N^n$.

Finally, 1000 Chebyshev nodes (corresponding to $N = 999$) were selected for computing the eigenvalue spectrum. The generalized eigenvalue problem was solved using the \texttt{eigs} function in the MATLAB environment.

\end{appen}\clearpage

\bibliographystyle{jfm}
\bibliography{jfm}

@article{Xie2018,
  author  = {Xie, Jin-Han and B{\"u}hler, Oliver},
  title   = {Exact third-order structure functions for two-dimensional turbulence},
  journal = {Journal of Fluid Mechanics},
  volume  = {851},
  pages   = {672--686},
  year    = {2018},
  doi     = {10.1017/jfm.2018.528}
}

@article{Xie2022,
  author  = {Xie, Jin-Han and Huang, Shi-Di},
  title   = {Bolgiano--Obukhov scaling in two-dimensional isotropic convection},
  journal = {Journal of Fluid Mechanics},
  volume  = {942},
  year    = {2022},
  doi     = {10.1017/jfm.2022.373}
}

@article{Zhu2023,
  author  = {Zhu, Hang-Yu and Xie, Jin-Han and Xia, Ke-Qing},
  title   = {Circulation in quasi-2D turbulence: Experimental observation of the area rule and bifractality},
  journal = {Physical Review Letters},
  volume  = {130},
  number  = {21},
  pages   = {214001},
  year    = {2023},
  doi     = {10.1103/PhysRevLett.130.214001}
}

@article{Wu2016,
  author  = {Wu, X. and Dong, M.},
  title   = {A local scattering theory for the effects of isolated roughness on boundary-layer instability and transition: transmission coefficient as an eigenvalue},
  journal = {Journal of Fluid Mechanics},
  volume  = {794},
  pages   = {68--108},
  year    = {2016},
  doi     = {10.1017/jfm.2016.125}
}

@article{Dong2021,
  author  = {Dong, M. and Zhao, L.},
  title   = {An asymptotic theory of the roughness impact on inviscid {Mack} modes in supersonic/hypersonic boundary layers},
  journal = {Journal of Fluid Mechanics},
  volume  = {913},
  pages   = {A22},
  year    = {2021},
  doi     = {10.1017/jfm.2021.12}
}

@article{Dong2022,
  author  = {Dong, M. and Zhang, M.},
  title   = {Asymptotic study of linear instability in a viscoelastic pipe flow},
  journal = {Journal of Fluid Mechanics},
  volume  = {935},
  pages   = {A28},
  year    = {2022},
  doi     = {10.1017/jfm.2022.48}
}

@article{Chen2019,
  author  = {Chen, X. and Huang, G. L. and Lee, C. B.},
  title   = {Hypersonic boundary layer transition on a concave wall: stationary {G\"ortler} vortices},
  journal = {Journal of Fluid Mechanics},
  volume  = {865},
  pages   = {1--40},
  year    = {2019},
  doi     = {10.1017/jfm.2019.55}
}

@article{Chen2022,
  author  = {Chen, X. and Dong, S. and Tu, G. and Yuan, X. and Chen, J.},
  title   = {Boundary layer transition and linear modal instabilities of hypersonic flow over a lifting body},
  journal = {Journal of Fluid Mechanics},
  volume  = {938},
  pages   = {A8},
  year    = {2022},
  doi     = {10.1017/jfm.2022.123}
}

@article{Kellay1995,
    author = {Kellay, H. and Wu, X.-l. and Goldburg, W. I.},
    title = {Experiments with Turbulent Soap Films},
    journal = {Phys. Rev. Lett.},
    volume = {74},
    issue = {20},
    pages = {3975--3978},
    year = {1995},
    doi = {10.1103/PhysRevLett.74.3975},
    publisher = {American Physical Society}
}

@article{Kellay2002,
    author = {Kellay, H. and Goldburg, W.},
    title = {Two-dimensional turbulence: A review of some recent experiments},
    journal = {Rep. Prog. Phys.},
    volume = {65},
    issue = {5},
    pages = {845},
    year = {2002},
    doi = {10.1088/0034-4885/65/5/204}
}

@article{Tran2010,
    author = {Tran, T. and Chakraborty, P. and Guttenberg, N. and Prescott, A. and Kellay, H. and Goldburg, W. and Goldenfeld, N. and Gioia, G.},
    title = {Macroscopic effects of the spectral structure in turbulent flows},
    journal = {Nat. Phys.},
    volume = {6},
    issue = {6},
    pages = {438--441},
    year = {2010},
    doi = {10.1038/nphys1608}
}

@article{Vilquin2021,
    author = {Vilquin, A. and Jagielka, J. and Djambov, S. and Herouard, H. and Fischer, P. and Bruneau, C.-H. and Chakraborty, P. and Gioia, G. and Kellay, H.},
    title = {Asymptotic turbulent friction in 2D rough-walled flows},
    journal = {Sci. Adv.},
    volume = {7},
    issue = {23},
    pages = {eabc6234},
    year = {2021},
    doi = {10.1126/sciadv.abc6234}
}

@article{Lvov2009,
    author = {L'vov, V. S. and Procaccia, I. and Rudenko, O.},
    title = {Velocity and energy profiles in two- versus three-dimensional channels: Effects of an inverse- versus a direct-energy cascade},
    journal = {Phys. Rev. E},
    volume = {79},
    issue = {4},
    pages = {045304(R)},
    year = {2009},
    doi = {10.1103/PhysRevE.79.045304}
}

@article{Samanta2014,
    author = {Samanta, D. and Ingremeau, F. and Cerbus, R. and Tran, T. and Goldburg, W. I. and Chakraborty, P. and Kellay, H.},
    title = {Scaling of near-wall flows in quasi-two-dimensional turbulent channels},
    journal = {Phys. Rev. Lett.},
    volume = {113},
    issue = {2},
    pages = {024504},
    year = {2014},
    doi = {10.1103/PhysRevLett.113.024504}
}

@article{Markeviciute2021,
    author = {Markeviciute, V. K. and Kerswell, R. R.},
    title = {Degeneracy of turbulent states in two-dimensional channel flow},
    journal = {J. Fluid Mech.},
    volume = {917},
    pages = {A57},
    year = {2021},
    doi = {10.1017/jfm.2021.262}
}

@article{Falkovich2018,
    author = {Falkovich, G. and Vladimirova, N.},
    title = {Turbulence appearance and nonappearance in thin fluid layers},
    journal = {Phys. Rev. Lett.},
    volume = {121},
    issue = {16},
    pages = {164501},
    year = {2018},
    doi = {10.1103/PhysRevLett.121.164501}
}

@article{tran2010macroscopic,
  title={Macroscopic effects of the spectral structure in turbulent flows},
  author={Tran, Tuan and Chakraborty, Pinaki and Guttenberg, Nicholas and Prescott, Alisia and Kellay, Hamid and Goldburg, Walter and Goldenfeld, Nigel and Gioia, Gustavo},
  journal={Nature Physics},
  volume={6},
  number={6},
  pages={438--441},
  year={2010},
  publisher={Nature Publishing Group UK London}
}

@article{kellay2012testing,
  title={Testing a missing spectral link in turbulence},
  author={Kellay, Hamid and Tran, Tuan and Goldburg, Walter and Goldenfeld, Nigel and Gioia, Gustavo and Chakraborty, Pinaki},
  journal={Physical review letters},
  volume={109},
  number={25},
  pages={254502},
  year={2012},
  publisher={APS}
}

@techreport{Kraichnan1967,
  title={Inertial ranges in two-dimensional turbulence},
  author={Kraichnan, Robert H},
  year={1967}
}

@article{Batchelor1969,
  title={Computation of the energy spectrum in homogeneous two-dimensional turbulence},
  author={Batchelor, George K},
  journal={Physics of Fluids},
  volume={12},
  number={12},
  pages={II--233},
  year={1969},
  publisher={AIP Publishing}
}

@article{xie2021low,
  title={A low-communication-overhead parallel DNS method for the 3D incompressible wall turbulence},
  author={Xie, Jiabin and He, Jianchao and Bao, Yun and Chen, Xi},
  journal={International Journal of Computational Fluid Dynamics},
  volume={35},
  number={6},
  pages={413--432},
  year={2021},
  publisher={Taylor \& Francis}
}

@book{1,
  title={Stability and transition in shear flows},
  author={Schmid, Peter J and Henningson, Dan S},
  volume={142},
  year={2012},
  publisher={Springer Science \& Business Media}
}

@techreport{2,
  author       = {Tollmien, W.},
  title        = {The production of turbulence},
  institution  = {United States. National Advisory Committee for Aeronautics},
  year         = {1931},
  number       = {609},
  series       = {NACA Technical Memorandum},
  address      = {Washington, D.C.},
  month        = mar,
  language     = {english},
  note         = {English translation of German original "Uber die entstehung der turbulenz" (Nachr. Ges. Wiss. Göttingen, Math. Phys. Klasse, 1929)}
}

@article{3,
author = {Schlichting, H.},
journal = {Nachrichten von der Gesellschaft der Wissenschaften zu Göttingen, Mathematisch-Physikalische Klasse},
pages = {181-208},
title = {Zur Enstehung der Turbulenz bei der Plattenströmung},
url = {http://eudml.org/doc/59420},
volume = {1933},
year = {1933},
}

@article{4,
    author = "Landau, Lev Davidovich",
    editor = "ter Haar, D.",
    title = "{On the Problem of Turbulence}",
    doi = "10.1016/b978-0-08-010586-4.50057-2",
    journal = "Compt. Rend. Acad. Sci. URSS",
    volume = "44",
    year = "1944"
}

@article{5,
    author = {Herbert, Thorwald},
    title = {Secondary instability of plane channel flow to subharmonic three‐dimensional disturbances},
    journal = {Physics of Fluids},
    volume = {26},
    number = {4},
    pages = {871-874},
    year = {1983},
    month = {04},
}

@article{6,
  author={Stuart, J. T.},
  title={On the non-linear mechanics of hydrodynamic stability},
  journal={Journal of Fluid Mechanics},
  volume={4},
  number={1},
  pages={1--21},
  year={1958},
  doi={10.1017/S0022112058000276}
}

@article{7, title={On the non-linear mechanics of wave disturbances in stable and unstable parallel flows Part 1. The basic behaviour in plane Poiseuille flow}, volume={9}, DOI={10.1017/S002211206000116X}, number={3}, journal={Journal of Fluid Mechanics}, author={Stuart, J. T.}, year={1960}, pages={353–370}}

@article{8, title={On the non-linear mechanics of wave disturbances in stable and unstable parallel flows Part 2. The development of a solution for plane Poiseuille flow and for plane Couette flow}, volume={9}, DOI={10.1017/S0022112060001171}, number={3}, journal={Journal of Fluid Mechanics}, author={Watson, J.}, year={1960}, pages={371–389}}

@article{9, title={Three-dimensional finite-amplitude solutions in plane Couette flow: bifurcation from infinity}, volume={217}, DOI={10.1017/S0022112090000829}, journal={Journal of Fluid Mechanics}, author={Nagata, M.}, year={1990}, pages={519–527}}

@article{10, title={Bifurcations in Couette flow between almost corotating cylinders}, volume={169}, DOI={10.1017/S0022112086000605}, journal={Journal of Fluid Mechanics}, author={Nagata, M.}, year={1986}, pages={229–250}}

@article{12,
  title={Finite amplitude stability of plane parallel flows},
  author={Herbert, Thorwald},
  journal={In AGARD Laminar-Turbulent Transition 10 p (SEE N78-14316 05-34},
  year={1977}
}

@article{13, title={Three-dimensional wavelike equilibrium states in plane Poiseuille flow}, volume={228}, DOI={10.1017/S0022112091002653}, journal={Journal of Fluid Mechanics}, author={Ehrenstein, U. and Koch, W.}, year={1991}, pages={111–148}}

@article{14, title={Coherent structures in wall-bounded turbulence}, volume={842}, DOI={10.1017/jfm.2018.144}, journal={Journal of Fluid Mechanics}, author={Jiménez, Javier}, year={2018}, pages={P1}}

@article{15,
  title={Exact coherent states and the nonlinear dynamics of wall-bounded turbulent flows},
  author={Graham, Michael D and Floryan, Daniel},
  journal={Annual Review of Fluid Mechanics},
  volume={53},
  number={1},
  pages={227--253},
  year={2021},
  publisher={Annual Reviews}
}

@article{17,
  title={Onset of transition in boundary layers},
  author={Herbert, Thorwald},
  journal={International journal for numerical methods in fluids},
  volume={8},
  number={10},
  pages={1151--1164},
  year={1988},
  publisher={Wiley Online Library}
}

@article{18,
  title={Secondary instability of boundary layers},
  author={Herbert, Thorwald},
  journal={Annual review of fluid mechanics},
  volume={20},
  pages={487--526},
  year={1988}
}

@article{19,
 ISSN = {09501207},
 URL = {http://www.jstor.org/stable/96239},
 author = {H. B. Squire},
 journal = {Proceedings of the Royal Society of London. Series A, Containing Papers of a Mathematical and Physical Character},
 number = {847},
 pages = {621--628},
 publisher = {The Royal Society},
 title = {On the Stability for Three-Dimensional Disturbances of Viscous Fluid Flow between Parallel Walls},
 urldate = {2025-08-22},
 volume = {142},
 year = {1933}
}

@article{21, title={Transition to turbulence in two-dimensional Poiseuille flow}, volume={218}, DOI={10.1017/S0022112090001008}, journal={Journal of Fluid Mechanics}, author={Jiménez, Javier}, year={1990}, pages={265–297}}

@article{22,
  title={Sur les équations différentielles linéaires à coefficients périodiques},
  author={Floquet, G},
  journal={Annales scientifiques de l'École Normale Supérieure, Serie 2},
  volume={12},
  pages={47--88},
  year={1883},
}

@article{23,
  author  = {Han, Yufeng and Liu, Jianxin and Luo, Jisheng},
  title   = {Improvement for expansion of parabolized stability equation method in boundary layer stability analysis},
  journal = {Applied Mathematics and Mechanics (English Edition)},
  volume  = {39},
  number  = {12},
  pages   = {1737--1753},
  year    = {2018},
  doi     = {10.1007/s10483-019-2401-9},
  url     = {https://doi.org/10.1007/s10483-019-2401-9}
}

@article{24,
  title={Secondary subharmonic instability of hypersonic boundary layer in thermochemical equilibrium over a flat plate},
  author={Kumar, Chandan and Prakash, Akshay},
  journal={Physics of Fluids},
  volume={33},
  number={2},
  year={2021},
  publisher={AIP Publishing}
}

@article{SVD,
  title={The approximation of one matrix by another of lower rank},
  author={Eckart, Carl and Young, Gale},
  journal={Psychometrika},
  volume={1},
  number={3},
  pages={211--218},
  year={1936},
  publisher={Springer-Verlag}
}

@article{orszag1980,
  title={Transition to turbulence in plane Poiseuille and plane Couette flow},
  author={Orszag, Steven A and Kells, Lawrence C},
  journal={Journal of Fluid Mechanics},
  volume={96},
  number={1},
  pages={159--205},
  year={1980},
  publisher={Cambridge University Press}
}

@article{kraichnan1980,
  title   = {Two-dimensional turbulence},
  author  = {Kraichnan, Robert H and Montgomery, David},
  journal = {Reports on Progress in Physics},
  volume  = {43},
  number  = {5},
  pages   = {547--619},
  year    = {1980}
}

@article{boffetta2012,
  title={Two-dimensional turbulence},
  author={Boffetta, Guido and Ecke, Robert E},
  journal={Annual review of fluid mechanics},
  volume={44},
  number={1},
  pages={427--451},
  year={2012},
  publisher={Annual Reviews}
}

@article{sobey1986,
  title={Bifurcations of two-dimensional channel flows},
  author={Sobey, Ian J and Drazin, Philip G},
  journal={Journal of Fluid Mechanics},
  volume={171},
  pages={263--287},
  year={1986},
  publisher={Cambridge University Press}
}

@article{chen2024,
  title={Unbounded two-dimensional wall turbulence induced by inverse cascade},
  author={Chen, Xi and Duan, Peng-Yu and He, Jianchao},
  journal={Physical Review Fluids},
  volume={9},
  number={3},
  pages={034609},
  year={2024},
  publisher={APS}
}

@article{kerswell2018,
  title={Nonlinear nonmodal stability theory},
  author={Kerswell, Rich R},
  journal={Annual Review of Fluid Mechanics},
  volume={50},
  pages={319--345},
  year={2018},
  publisher={Annual Reviews}
}

@article{jimenez1996,
  title={The structure of the vortices in freely decaying two-dimensional turbulence},
  author={Jim{\'e}nez, Javier and Moffatt, HK and Vasco, Carlos},
  journal={Journal of Fluid Mechanics},
  volume={313},
  pages={209--222},
  year={1996},
  publisher={Cambridge University Press}
}

@article{jimenez2020,
  title={Dipoles and streams in two-dimensional turbulence},
  author={Jim{\'e}nez, Javier},
  journal={Journal of Fluid Mechanics},
  volume={904},
  pages={A39},
  year={2020},
  publisher={Cambridge University Press}
}

\end{document}